\documentclass[a4paper,11pt]{article}
\usepackage{jheppub} 
\usepackage{lineno}
\usepackage[table, dvipsnames]{xcolor}
\usepackage{hyperref}
\usepackage{graphicx}
\usepackage{subfig}
\usepackage{amsmath,amssymb,amsthm,amsfonts}
\usepackage{mathtools}
\usepackage[all]{xy}
\usepackage{tikz}
\usepackage{float}
\usetikzlibrary{calc}
\usepackage[toc,page]{appendix}
\usepackage{enumerate}
\usepackage{booktabs}
\usepackage{cancel}
\usepackage{setspace}
\usepackage{bbm}
\usepackage{youngtab}
\usepackage{slashed}
\usepackage{fancyhdr}
\usepackage{empheq}
\usepackage{multirow}
\usepackage{makecell}
\usepackage{float}
\usepackage{comment}
\usepackage{tensor}
\usepackage{subcaption}
\usepackage{verbatim}
 \usepackage[framemethod=tikz]{mdframed}
\usepackage{soul}
\usepackage{cancel}
\usepackage[normalem]{ulem}
\usepackage{mathtools}
\usepackage{physics}
\usepackage{braket}
\usepackage{scalerel}
\usepackage{array}
\usepackage{enumitem}
\usepackage{tcolorbox}
\usepackage[backgroundcolor=white!95!red, linecolor=purple, bordercolor=purple, textsize=footnotesize]{todonotes}

\newcommand{\ab}{|}

\newcommand{\vol}{\text{vol}}

\newcommand{\Int}{\text{int}}
\newtheorem{nogo}{dS No-go}
\renewcommand{\arraystretch}{1.7}
\definecolor{Gray}{gray}{0.9}
\definecolor{mycolor}{RGB}{240,240,240}

\title{\boldmath On (A)dS Solutions from Scherk-Schwarz Orbifolds}

\author{Susha Parameswaran and Marco Serra}
\affiliation{Department of Mathematical Sciences, University of Liverpool,\\
Liverpool, L69 7ZL, United Kingdom}

\emailAdd{susha@liv.ac.uk}
\emailAdd{Marco.Serra@liverpool.ac.uk}

\abstract{We investigate the existence of dS vacua in supersymmetry-breaking Scherk-Schwarz toroidal compactifications of type II string theory, using the well-understood ingredients of curvature, fluxes and 1-loop Casimir energy.  Starting from the 10d equations, we derive a series of no-go theorems and existence conditions for dS, and present  two explicit, fully-backreacted, solutions: a dS one, which turns out to be not under  control,  and an AdS one, which can be chosen at arbitrarily weak coupling and large volume by dialling the unbounded fluxes.  We then use a lower-dimensional EFT description to show that {\it{any}} dS solution has a universal tachyon and no parametric control.  The simplest AdS solutions are also perturbatively unstable.  We extend the no-go theorems to slow-roll acceleration and test various swampland conjectures in our non-supersymmetric string setup.  The question of numerically controlled, unstable dS is left open.}

\begin{document}
\maketitle

\flushbottom

\section{Introduction}

Do well-controlled, metastable de Sitter vacua exist in perturbative string theory?  Despite two decades of intense investigation \cite{Cicoli:2018kdo, Cicoli:2023opf}, there is still no consensus on the answer to this question, but it is generally expected that it will lead us to deep insights into the nature of quantum gravity and into the nature of the dark energy that currently dominates our Universe.  On the one hand, large numbers of de Sitter vacua would address the cosmological constant problem and explain dark energy \cite{Bousso:2000xa}; on the other hand it has been conjectured that metastable de Sitter vacua may not exist at all \cite{Danielsson:2018ztv, Obied:2018sgi}, or may not exist at the asymptotics of field space where we have good control \cite{Ooguri:2018wrx, Bedroya:2019snp}.  The present paper studies this question a hitherto unsearched region of the string landscape, namely supersymmetry-breaking Scherk-Schwarz toroidal orbifold compactifications of type II string theory.  These constructions are very well understood, with a worldsheet description of the supersymmetry-breaking and subsequent Casimir energy, and -- as we will see -- the possibility of fully-backreacted ten-dimensional supergravity solutions with just the simple ingredients of fluxes, curvature, and Casimir.  Without a doubt, it should be established whether or not there are de Sitter spacetimes amongst these solutions.

In fact, there is good motivation to expect that this might be possible.  The famous Maldacena-Nu\~nez no-go theorem \cite{Maldacena:2000mw}, derived simply from the (integrated) equations of motion, rules out non-singular warped compactifications to $d$-dimensional de Sitter spacetime ($d\geq 2$) in $D$-dimensional two-derivative gravity theories ($D >2$) whose sources satisfy the strong energy condition, $\tilde{T} \equiv -T^\mu_\mu + \frac{d}{D-2}T^M_M \geq 0$ (where $\mu=0,1,\dots, d-1$ and $M=0,1,\dots, D-1$).  This has been extended, using a tree-level worldsheet analysis, to all orders in $\alpha'$: there are no classical de Sitter vacua with $d\geq 4$ in the heterotic string theory or in type II without RR fluxes \cite{Kutasov:2015eba}.  Sources satisfying the Maldacena-Nu\~nez strong energy condition assumption include fluxes and positive-tension $D$-branes, but not $O$-planes.  There has subsequently been an enormous amount of work dedicated to searching for classical de Sitter vacua in type II orientifolds with RR fluxes (see the review \cite{Andriot:2019wrs} and references therein).  One difficulty in these constructions is solving for the backreaction of the localised sources; generally solutions can be found only in the smeared approximation (see, however, \cite{Tomasiello:2022wfq} for interesting progress on this front). Irrespective of this challenge, all classical de Sitter solutions found thus far are perturbatively unstable \cite{Danielsson:2011au,Shiu:2011zt,Danielsson:2012et,Junghans:2016uvg, Andriot:2022bnb} and inconsistent with the weak coupling and/or large volume expansions \cite{Junghans:2018gdb}, \cite{Andriot:2024cct}.  At the same time, supersymmetry-breaking provides other ways to break the strong energy condition.

In the ten-dimensional non-supersymmetric tachyon-free string theories -- i.e. $SO(16)\times SO(16)$ heterotic, $U(32)$ type 0'B, and the $USp(32)$ brane supersymmetry-breaking Sugimoto model -- the string-scale supersymmetry-breaking leads to 10d dilaton-tadpole potentials that do indeed break the strong energy condition.  However, whilst AdS solutions can be found by balancing the runaway dilaton against fluxes \cite{Mourad:2016xbk, Baykara:2022cwj}, and Minkowski solutions can be obtained by allowing a non-trivial dilaton-profile \cite{Dudas:2000ff}, it turns out that the de Sitter no-go theorems can be straightforwardly extended to these non-supersymmetric theories \cite{Basile:2020mpt}.  Consistent with these extended no-go theorems, there are also intriguing results against the uplift to de Sitter of certain anti de Sitter vacua, $AdS_3 \times S^3 \times S^3 \times S^1$ with NSNS fluxes, in the non-supersymmetric $SO(16) \times SO(16)$ heterotic theory \cite{Baykara:2022cwj}.  Since there is a good worldsheet description of these vacua at tree-level, and moreover an exact computation of the 1-loop Casimir energy, this result may be seen as a hint towards a 1-loop generalisation of the tree-level all $\alpha'$ orders result in \cite{Kutasov:2015eba}. 

In Scherk-Schwarz toroidal orbifolds, the supersymmetry-breaking takes place spontaneously at the compactification scale.  A 1-loop effective potential, i.e the Casimir energy, is generated, which induces a runaway in both the dilaton and radion, and again breaks the strong energy condition.  In the following, we will consider how the runaway 1-loop Casimir potential can be balanced against tree-level curvature and flux terms, in such a way as to stabilise the moduli in a simple ``quantum de Sitter solution''. Casmir, curvature and fluxes were used in particular warped hyperbolic compactifications of M-theory, with a localised Casimir term supported by a rather non-trivial internal geometry, to stabilise the volume modulus in a de Sitter vacuum in \cite{DeLuca:2021pej}.  Within perturbative string theory, and with simple toroidal geometries, it may seem at first sight difficult to balance tree-level and 1-loop terms self-consistently at weak coupling and large volume without running into the Dine-Seiberg problem \cite{Dine:1985he}, since the 1-loop term will be suppressed with respect to the tree-level terms, both in $g_s$ and in the radius of the Scherk-Schwarz torus, $R_{\text{ss}}$.  However, in the case of product compactifications, $dS_{d} \times Y^m \times \mathcal{T}_{\text{ss}}^{10-d-m}$, the interplay between two compactification scales, say $R$ and $R_{\text{ss}}$, together with the string coupling, in principle allows the tree-level and 1-loop contributions to compete.  Moreover, the fluxes can be turned on in a way that does not induce any RR tadpoles; therefore no localised sources are necessary and flux numbers can be unbounded and dialled to assist with parametric control \cite{DeWolfe:2005uu, Junghans:2018gdb}.

With these motivations in mind, we will first derive a series of  no-go theorems for de Sitter vacua in Scherk-Schwarz toroidal orbifold compactifications of type II string theories, starting from the 10d supergravity equations of motion plus the 1-loop Casimir contribution (the latter computed via the worldsheet theory in the limit that volumes are large so that curvature and fluxes can be neglected).  These no-go theorems can be read as giving necessary ingredients for de Sitter solutions.  We then use them to identify an explicit example four-dimensional de Sitter solution that solves all the ten-dimensional equations of motion and Bianchi identities, alongside an example anti de Sitter solution.  We find, however, that the de Sitter solution is not under  control, whereas the anti de Sitter solution -- which is not scale separated -- can be chosen at arbitrarily weak coupling and large volume by tuning the unbounded flux numbers.  We then turn to the lower-dimensional EFT description of the compactifications, which can be applicable even in the absence of scale separation for consistent truncations.   We use this EFT to show that any putative de Sitter  solution has a universal tachyon (we do not find such general results for anti de Sitter solutions, but the anti de Sitter example that we present also turns out to be unstable, with a tachyon below the BF bound).  Moverover, we can use the de Sitter no-gos together with the EFT description to rule out any parametrically controlled de Sitter.  The EFT also rules out quasi-de Sitter ``slow-roll'' potentials.   

The paper is organised as follows. Section \ref{10dsol} provides a 10d analysis, where no-go theorems and necessary conditions for de Sitter are proven and explicit de Sitter and anti de Sitter solutions are derived.  Section \ref{ddimEFT} uses a lower-dimensional EFT analysis to draw general conclusions on the perturbative stability and parametric control of any (anti) de Sitter solution, as well as the possibility of ``slow-roll''; alongside this, we briefly test various swampland conjectures in our non-supersymmetric top-down setting.  We summarise our results and provide some outlook in section \ref{Conclusions}.  For convenience, we provide some technical details, on Scherk-Schwarz orbifolds and their 1-loop effective  potentials and on the 10d supergravity equations of motions, in the appendices, which we refer to in the main text.

\section{10d (anti) de Sitter solutions -- no-gos and existence conditions}\label{10dsol}

In this section we investigate the possibility of (anti) de Sitter solutions within 10d Scherk-Schwarz compactifications, including the 1-loop Casimir potential that arises due to supersymmetry breaking.  After a brief review of Scherk-Schwarz compactifications and the Casimir potential, we present our compactification ansatz. We then use a selection of the 10d equations of motion to derive a series of no-go theorems for de Sitter solutions, which may also be read as giving necessary ingredients for de Sitter.  Finally, we use these no-go theorems to identify two examples of concrete, fully-backreacted, non-supersymmetric solutions to all the 10d equations of motion and Bianchi identities; one de Sitter and one anti de Sitter.  We make some initial comments on their self-consistency within the supergravity approximations used, and explore this issue -- together with the perturbative stability properties of the solutions -- more generally in the next section.

\subsection{Scherk-Schwarz compactification and the 1-loop Casimir potential}
We consider unwarped compactifications of 10d type II strings given by the direct product
\begin{equation}\label{10comp}
\mathcal{M}_{1,d-1}\times \mathcal{C}^{10-d}
\, , 
\end{equation}  
 where $\mathcal{M}_{1,d-1}$ is a $d$-dimensional (anti) de Sitter spacetime and $\mathcal{C}^{10-d}$ is the internal compactification space. We are interested in two scenarios, namely a \textit{single-factor compactification}, where the internal space is identified with a flat   ``Scherk-Schwarz'' torus $\mathcal{T}^n_{\text{ss}}$
\begin{equation}\label{isocomp}
    \mathcal{C}^{10-d}=\mathcal{T}^n_{\text{ss}}
\end{equation}
and a more general \textit{product compactification}, where the internal space is taken to be an unwarped, direct product of a  Euclidean compact space $Y^m$, which can be curved, with $m=\text{dim}(Y)$, and the flat Scherk-Schwarz torus
\begin{equation}\label{anisocomp}
    \mathcal{C}^{10-d}=Y^m\times  \mathcal{T}^n_{\text{ss}} \,.
\end{equation}
 In both scenarios, the torus $\mathcal{T}^n_{\text{ss}}$ is orbifolded with the $\mathbb{Z}_2$ generator $g=(-1)^F\delta_{KK}$, where $F$ is the spacetime fermion number and $\delta_{KK}$ the freely-acting shift $x^i_L\rightarrow x^i_L+\pi R_{\text{ss}}\sqrt{\alpha'}/2$, $x^i_R\rightarrow x^i_R+\pi R_{\text{ss}}\sqrt{\alpha'}/2$ on the torus coordinates $x^i=x^i_L+x^i_R$, $i=d+m,\dots,9$ (see Appendix \ref{ssappendix} and also  \cite{Angelantonj:2002ct} for review).
For concreteness, we assume an isotropic, factorisable torus
 \begin{equation} \label{torusfact}
     \mathcal{T}^n_{\text{ss}}=\prod_{i=d+m}^{9} S^1_i(R_{\text{ss}})\, ,
 \end{equation}
 with $R_{\text{ss}}$ the ``Scherk-Schwarz radius'' in units of $\sqrt{\alpha'}$. The effect of this orbifold is
 to realise in string-theory a spontaneous supersymmetry-breaking through a coordinate-dependent compactification \`a la Scherk-Schwarz \cite{Scherk:1978ta} (for a more recent review see \cite{Antoniadis:2023doq}), where antiperiodic boundary conditions are implemented for the fermionic fields along the $n$ circles in the torus.  
Indeed, $g$ acts on the torus lattice by shifting the KK number of the fermions, such that all the zero modes of the fermions, gravitini included, acquire a tree-level mass 
\begin{equation}
{M_{\text{ss}}\sim \frac{M_s}{R_{\text{ss}}}}
\end{equation}
(with $M_s=1/\sqrt{\alpha'}$ the string scale).  Supersymmetry-breaking is thus set at the  Kaluza-Klein scale associated to $\mathcal{T}^{n}_{\text{ss}}$.  At the same time, modular invariance introduces into the spectrum twisted states   associated to a ``wrong'' GSO projection. These states come with non-trivial winding masses and include a scalar that becomes tachyonic at the Hadgedorn radius $R_H=\sqrt{2}$, where the 1-loop effective potential, generated by the supersymmetry-breaking, diverges.  We are therefore  interested in the regime with large internal radius, ${R_{\text{ss}}}\gg 1$, for which the supersymmetry-breaking scale is much lower than the string scale, $M_{\text{ss}}\ll M_s$. With this assumption, we are protected from the Hadgedorn instability, the 1-loop potential is finite and all the twisted states, being very massive, can be integrated out from the EFT. 

The 1-loop effective potential of the lower $(10-n)$-dimensional theory can be computed exactly in the 10d string-frame at the worldsheet level by integrating the torus partition function $\mathcal{T}$  on the $SL(2,\mathbb{Z})$ fundamental domain $\mathcal{F}$, parametrized by the Teichmuller parameter $\tau$. In the limit $R_{\text{ss}}\gg 1$, we obtain (see Appendix \ref{one-loopotA})
\begin{equation} \label{casimirpot}
\begin{split}
    V_{\text{ss}}(R_{\text{ss}})&\equiv -\frac{M_s^{10-n}}{(2\pi)^{10-n}}\int_\mathcal{F}\frac{d^2\tau}{2(\text{Im}\,\tau)^2}\mathcal{T}\\
    &=(n_f^{0}-n_b^{0})\,{\xi_n}\,M_{\text{ss}}^{10-n}+\mathcal{O}\left(\,e^{-\frac{M_s}{M_{\text{ss}}}}\right)\,,
 \end{split}   
\end{equation}
 where $\xi_n=\frac{3\,\cdot\, 2^{3n-7}}{\pi^{15-n}}\,\frac{1}{n^5}$ is  an $n$-dependent dressing factor (computed in (\ref{xin})) and $n_f^{0}$ ($n^{0}_b$) the number of massless fermions (bosons) with $n_f^{0}-n_b^0=-64$. The leading-order contribution, set by the KK modes associated to the massless string oscillators propagating along the supersymmetry-breaking directions, has a negative runaway behavior that pushes the Scherk-Schwarz radius towards the tachyonic regime; all the remaining states, i.e. massive string states and KK modes thereof, winding modes and the non-level matched states, are very massive  and yield exponentially suppressed contributions, which can be safely neglected.

From a 10d point of view, this 1-loop potential can be thought as originating from  a 1-loop Casimir-like energy density $\hat{V}_{\text{ss}}$, which should be distributed evenly throughout the 10d spacetime: 
\begin{equation}
    \hat{V}_{\text{ss}}(R_{\text{ss}})=(n_f^{0}-n_b^{0}){\xi_n}\frac{1}{(2\pi)^n}\left(\frac{M_s}{R_{\text{ss}}}\right)^{10}\, .
\end{equation}
The 10d 1-loop potential $\hat{V}_{\text{ss}}$  then leads to a 1-loop contribution to the 10d stress-energy tensor, which we denote as $T^{\text{ss}}_{MN}$ ($M,N=0,1,\dots, 9$).  It acquires the following form (see Appendix \ref{ssappendix}, also cfr. Appendix A \cite{Arkani-Hamed:2007ryu}, (4.2) \cite{DeLuca:2021pej})
\begin{equation}\label{tsb}
        T^{\text{ss}}_{MN}=-\hat{V}_{\text{ss}}\left(g_{MN}-\,10\,\frac{1}{n}\,g_{ij}\,\delta^i_M\delta^j_N\right)\,,
\end{equation}
with $i,j = 10-n, \dots, 9$ indices on $\mathcal{T}^n_{ss}$. Notice that $T_{MN}^{\text{ss}}$ is 10d-traceless\footnote{This result is actually independent of the simplifying assumption made here that the torus is factorisable and each circle in the torus has the same radius, $R_{\text{ss}}$, as can be seen from (\ref{TMNgentor}).}, indeed, denoting $T_{(10)}^{\text{ss}} \equiv g^{MN}T^{\text{ss}}_{MN}$
\begin{equation}\label{tracetcas}
   T^{\text{ss}}_{(10)}= -10\hat{V}_{\text{ss}}+\frac{1}{n}10\,n\hat{V}_{\text{ss}}=0\, .
\end{equation}
Consequently, the \textit{Strong Energy Condition} (SEC) of \cite{Maldacena:2000mw} that forbids $d$-dimensional de-Sitter spacetime 
\begin{equation}\label{sec}
    \tilde{T}^{ss}\equiv-T_{(d)}^{\text{ss}}+\frac{d}{8}T^{\text{ss}}_{(10)}\ge 0
\end{equation}
\textit{can be violated} if\footnote{Note the difference with the  ``dilaton tadpole potentials'' $V_{\text{10d}}(\Phi)$ in the tachyon-free 10d non-supersymmetric string theories, where $T_{MN}^{\text{10d non-susy}}=-V_{\text{10d}}(\Phi)\,g_{MN}$ and thus $\tilde{T}^{\text{10d non-susy}} \geq 0$ is violated for $V_{\text{10d}}(\Phi)>0$.} $T^{\text{ss}}_{(d)}\equiv g^{\mu\nu}T^{\text{ss}}_{\mu\nu}>0$ ($\mu=0,\dots, d-1$), i.e if $\hat{V}_{\text{ss}}<0$, that is if $n_b^{0}>n_f^{0}$. A negative Casimir energy sourced by a surplus of massless bosons is indeed the case for type II Scherk-Schwarz toroidal orbifolds since all the bulk fermions have been massed up by the antiperiodic boundary conditions, $n_f^{0}=0$, leaving us with the $n_b^{0}=64$ massless bosons from the closed-string spectrum.

Having seen that a negative 1-loop Casimir energy might help to obtain de Sitter, without the need  to introduce localised sources such as $O$-planes, we now investigate under which conditions the Scherk-Schwarz runaway can be stabilised in a (quantum) de Sitter solution  within type II flux compactifications.

\subsection{Compactification ansatz}

Let us first present the class of solutions that we consider. For the background (\ref{10comp})-(\ref{anisocomp}) 
 we write down the 10d string-frame metric as  
\begin{equation}\label{10ans}    ds^2_{10}=g^{S}_{\mu\nu}dx^\mu dx^\nu+\,g_{ab}\,dy^ady^b+\, g_{ij}\,dy^{i}dy^{j}\, ,
\end{equation}
 where we have split the 10d latin index $M=\{\mu,a,i\}$ in terms of  indices, respectively, on $\mathcal{M}_{1,d-1}$, $Y^m$ and $\mathcal{T}^n_{ss}$ and the $g_{ab}$ factor is understood to be absent in the  single-factor compactification case.  The torus is made compact via the identification $y^i\sim y^i+\ell_s$, with $\ell_s:=2\pi\sqrt{\alpha'}$. The superscript ``$S$'' on $g_{\mu\nu}^S$ reminds us that in this section we are working in the 10d string-frame, whilst in the next section the $d$-dimensional Einstein-frame will be used for the effective field theory description.

 Since we are mainly interested in de Sitter \textit{vacuum solutions}, we set the 10d dilaton $\Phi$  to its vev $\Phi=\Phi_0= \text{Log}(g_s)$.   We can thus neglect all the dilaton kinetic terms  in the equations. 
 We remark that,  in the absence of localised sources, a constant dilaton and a vanishing warp factor in (\ref{10ans}) can  solve the equations of motion without the need of any smearing approximation. In this respect,  once the backreaction of fluxes is taken into account, the solutions to the 10d equations of motion will be fully-backreacted.

Turning to the flux conventions, we decompose each of the NSNS and RR $q$-form field strengths, here denoted loosely as $F_q$,  into an internal component, $F_q^{\text{int}}$, threading the compact directions and an external one, $F_q^{\text{ext}}$, extending throughout the $d$-dimensional spacetime.  Requiring maximal symmetry in the spacetime forces the external component of the fluxes to be spacetime-filling; thus $F_{q}^{\text{ext}}$
can be present only if $d\le q$. We work in the ``magnetic-frame'' convention where $F_q^{\text{ext}}$ is expressed through its dual $F_{10-q}^{\text{int}}$, such that only internal fluxes will appear in the equations of motion. That is, we decompose a $q$-form flux $F_q$ as 
\begin{equation}\begin{split}
F_q&=F_q^{\text{ext}}+F_q^{\text{int}}\equiv (-1)^{[\frac{q+1}{2}]}\star_{10}F_{10-q}^{\text{int}}+F_q^{\text{int}}\\
&\qquad\qquad \qquad\,=(-1)^{[\frac{q+1}{2}]}(-1)^{(10-q)d}\vol_d\wedge \star_{10-d}F_{10-q}^{\text{int}}+F_q^{\text{int}}\, .
\end{split}
\end{equation}
It follows that
\begin{equation}
\ab F_q\ab^2=\ab F_q^{\text{int}}\ab^2+\ab F_q^{\text{ext}}\ab^2=\ab F_q\ab^2_\Int-\ab F_{10-q}\ab^2_\Int\, ,
\end{equation}
where the subscript $\ab \cdot \ab_\Int$ reminds us that the contractions are made using the internal positive-definite  metric. 
Therefore, besides the standard $F_1$, $F_3$ and $F_5$ fluxes in IIB and $F_0$, $F_2$ and $F_4$ fluxes in IIA and the common $H_3$ flux, we will also deal with $F_6$, $F_7$  and $H_7$ fluxes for the appropriate spacetime dimensions.

When dealing with the  product compactification, we further decompose the internal part of the field strengths $F_q^{\text{int}}$ in terms of wedge products of form components defined on the $Y^m$ and $\mathcal{T}_{\text{ss}}^n$ spaces. To this purpose we introduce  the flat vielbeins  $\{e^{\dot{a}},e^{\dot{i}}\}$ defined with respect to the internal metrics  as $g_{ab}dy^ady^b=\delta_{\dot{a}\dot{b}}e^{\dot{a}}e^{\dot{b}}$ and $g_{ij}dy^idy^j=\delta_{\dot{i}\dot{j}}e^{\dot{i}}e^{\dot{j}}$.  Any internal $q$-form, such as fluxes $F_q^{\text{int}}$, can then be decomposed accordingly as sum of individual components \cite{Andriot:2016xvq}, e.g
\begin{equation}\label{fluxintdeco}
\begin{split}
    F_{q}^\Int&=\frac{1}{q!}F^{(0)}_{\dot{i}_1\dot{i}_2\dots \dot{i}_q}e^{\dot{i}_1}\wedge e^{\dot{i}_2}\wedge\dots \wedge e^{\dot{i}_q}+\frac{1}{(q-1)!} F^{(1)}_{\dot{a}_1\dot{i}_2\dots \dot{i}_q}e^{\dot{a}_1}\wedge e^{\dot{i}_2}\wedge\dots \wedge e^{\dot{i}_q}+\dots\\
    &=\sum_{s_q=0}^m F_q^{\text{int}(s_q)}\,,
    \end{split}
\end{equation}
where $F_q^{(s_q)}$ indicates the $q$-flux component with  $s_q=\{0,1,\dots,q\}\le m$ legs parallel to the $Y^m$-space and $q-s_q \leq n$ legs parallel to the $\mathcal{T}^n_{\text{ss}}$ space. We denote with $p_3$ and $p_7$ the legs that respectively $H_3$ and $H_7$ have along $Y^m$, to differentiate them from $F_3$ and $F_7$'s $s_3$ and $s_7$. It then follows that
\begin{equation}\label{fluxdeco}
    \ab F_{q}\ab^{2}_\Int=\sum_{s_q=0}^q\ab F_{q}^{(s_q)}\ab^2_\Int\,\quad,\,\ab F_{q}^{(s_q)}\ab^2_\Int\equiv\frac{1}{s_q!(q-s_q)!}{F_q}_{a_1\dots a_{s_q}i_{s_q+1}\dots i_{q}}{F_q}^{a_1\dots a_{s_q}i_{s_q+1}\dots i_{q}}\, .
\end{equation}
 Below, the following identity will also be useful: 
\begin{equation}\label{sqfqid}
    \sum_{s_q=0}^q s_q\ab F_q^{(s_q)}\ab^2_\Int=\ab F_q\ab^2_\Int-\ab F_q^{(0)}\ab^2_\Int+\sum_{s_q\ge 2}^q(s_q-1)\ab F_q^{(s_q)}\ab^2_\Int\, .
\end{equation}

\subsection{No-go theorems and existence conditions}
\label{nogosexist}
With our ansatz in hand, we will now investigate the 10d equations of motion (eom) to establish under which conditions a de Sitter solution is possible.  In particular, in this section, we  work in the 10d string-frame and consider  the 10d dilaton equation and the traces of the 10d trace-reversed Einstein equations.  This will be sufficient to formulate a series of {\it{no-go theorems for de Sitter solutions}}.  At the same time, we identify necessary ingredients for de Sitter solutions, and anti de Sitter solutions, for which one must also check the remaining components of the Einstein equations (\ref{10dEinstein}), the flux equations of motion (\ref{fluxeom}) and (\ref{h3eom}), and the Bianchi identities  (\ref{fluxbi}).  We do this for a candidate de Sitter solution and a candidate anti de Sitter solution in the following subsection.  Finally, the self-consistency of these solutions with the weak coupling and large volume expansions that have been assumed should be checked; we make this analysis in the following section.

The first equation that we consider is the
  10d dilaton eom. At tree-level, this equation is given in (\ref{10ddilatoneq}).  Since the Casimir term is a 1-loop effect, it does not couple to the 10d dilaton in the string-frame; the dilaton dynamics is therefore still governed by the classical equation (\ref{10ddilatoneq}), even in the presence of the Casimir term. For a constant dilaton it reads 
\begin{equation}\label{10ddilateqsm}
\begin{split}
    2\mathcal{R}_{10}=\ab H_3\ab^2_{\text{int}}-\ab H_7\ab^2_{\text{int}}\,.
\end{split}
\end{equation}

We now pass to the (trace-reversed) 10d Einstein equations
\begin{equation}\label{tracerevloop}
    \mathcal{R}_{MN}=\kappa_{10}^2 g_s^2\,\left( T_{MN}-\frac{1}{8}g_{MN}T_{(10)}\right)\, ,
\end{equation}
where the total stress-energy tensor, $T_{MN}$, is the sum of the classical contribution, $T^\text{SUGRA}_{MN}$ (\ref{10dTsugra}), and the 1-loop Casimir contribution, $T^{\text{ss}}_{MN}$ (\ref{tsb})
\begin{equation}
T_{MN}=T^{\text{SUGRA,IIA/B}}_{MN}+T^{\text{ss}}_{MN}\, .
\end{equation}

To start with, let us  extract the 10d trace. Since $T^{\text{ss}}_{MN}$ is traceless (\ref{tracetcas}), it does not contribute and the classical equation receives no corrections.  It can be written for IIA and IIB in the same fashion as\footnote{It is understood $q$  even (odd) for IIA (IIB).}
\begin{equation}\label{10dtraceeqsm}
   \begin{split}
\mathcal{R}_{10}=\frac{1}{4}\ab H_3\ab_{\text{int}}^2-\frac{1}{4}\ab H_7\ab_{\text{int}}^2+\frac{g_s^2}{8}\sum_{q=0}^{10-d}(5-q)\ab F_q\ab_{\text{int}}^2\, .
\end{split}
\end{equation}
  We can also extract from  (\ref{tracerevloop}) the $d$-dimensional trace equation
\begin{equation}\label{dtraceeqsm}
\begin{split}
\mathcal{R}_{d}=\frac{d}{16}\left(-\left(6\ab H_{7}\ab^2_{\Int}+2\ab H_{3}\ab^2_{\Int}\right)-g_s^2\sum_{q=0}^{10-d}(q-1)\ab F_{q}\ab^2_{\Int}+\frac{16}{d}\kappa_{10}^2g_s^2 T^{\text{ss}}_{(d)}\right)\, .
   \end{split}
\end{equation}
So far we obtained the dilaton equation of motion (\ref{10ddilateqsm}), the 10d Einstein trace (\ref{10dtraceeqsm}), and the $d$-dimensional Einstein trace (\ref{dtraceeqsm}). We now manipulate them.

As we have observed, the Casimir term is absent in both the 10d dilaton eq. (\ref{10ddilateqsm}) and 10d Einstein trace eq. (\ref{10dtraceeqsm}).  Moreover,  by suitable  linear combinations of these two equations we can eliminate either $\mathcal{R}_{10}$ or the NSNS fluxes. Taking the linear combination $(\ref{10ddilateqsm})-2\times(\ref{10dtraceeqsm})$ we eliminate $\mathcal{R}_{10}$ to obtain the following constraint equation
\begin{equation}\label{constraintsm}
    2(\ab H_3\ab_{\Int}^2-\ab H_7\ab_{\Int}^2)-g_s^2\sum_{q=0}^{10-d}(5-q)\ab F_q\ab^2_{\Int}=0\,.
\end{equation}
We can  subsequently eliminate $\ab H_3\ab^2_\Int$ 
in the $d$-dimensional Einstein trace to obtain an extension of the Maldacena-Nunez no-go theorem \cite{Maldacena:2000mw} in the presence of the 1-loop Casimir term:
\begin{equation}\label{dtraceh7}
\boxed{\mathcal{R}_{d}=\frac{d}{4}\left(-2\ab H_7\ab^2_{\text{int}} -g_s^2\sum_{q=0}^7\ab F_q\ab^2_\Int+\frac{4}{d}\kappa^2_{10}g_s^2T^{ss}_{(d)}\right)}
\end{equation}
We conclude
\begin{nogo}\label{signcas}
    There is no de Sitter solution if $T^{\text{ss}}_{(d)}<0\,$, that is if $n_f\ge n_b$.
\end{nogo}
\noindent Hence, a necessary (not sufficient) condition for a de Sitter solution is
$T^{\text{ss}}_{(d)}>0$, which is the case of a negative Casimir energy originating from a surplus of massless bosons. Indeed, such a Casimir energy \textit{violates} the Strong Energy Condition (\ref{sec}) as we have anticipated already.

As a next step,  it is useful to express $\mathcal{R}_{d}$ in terms of fluxes and the internal curvature $\mathcal{R}_{10-d}$. To do so, we consider another linear combination $(\ref{10dtraceeqsm})$-$\frac{1}{4}(\ref{10ddilateqsm})$, which eliminates  the NSNS fluxes, and use $\mathcal{R}_{10} = \mathcal{R}_{d}+\mathcal{R}_{10-d}$ to  obtain
\begin{equation}\label{rdrint}
\begin{split}
    \mathcal{R}_{d}&=-\mathcal{R}_{10-d}+\frac{1}{4}g_s^2\sum_q(5-q)\ab F_q\ab^2_{\Int} \,.
 \end{split}   
\end{equation}
We now continue our analysis by considering the single-factor and product compactifications in turn.

\subsubsection{Single-factor compactification}\label{isocompsec}
In this case, the internal space $\mathcal{C}^{10-d}=\mathcal{T}^{10-d}_{\text{ss}}$ is the Scherk-Schwarz torus and therefore the  internal curvature, $\mathcal{R}_{10-d}=\mathcal{R}_{10}-\mathcal{R}_d$, vanishes. Combining (\ref{10dtraceeqsm}) with (\ref{dtraceh7}) and (\ref{constraintsm}), we obtain the Einstein trace equation for the vanishing internal curvature: 
\begin{equation}\label{intcurviso}
   0=\frac{3}{2}\ab H_7\ab^2_\Int\delta_{d,3}+\frac{1}{4}g_s^2\sum_q(d+5-q)\ab F_q\ab^2_\Int-g_s^2\,\kappa^2_{10}\, T^{\text{ss}}_{(d)}\, .
\end{equation}
Let us notice that the Scherk-Schwarz contribution enters eqs (\ref{dtraceh7}) and (\ref{intcurviso}) with opposite signs: assuming $T^{\text{ss}}_{(d)}>0$ to avoid \textbf{dS No-go} \ref{signcas}, it contributes positively to the spacetime curvature but negatively to the internal one. Hence a compactification on a flat space can be achieved by cancelling  the (semi-)positive energy density from fluxes in (\ref{intcurviso}) with the negative one from the Scherk-Schwarz term. 
Then, from (\ref{rdrint}), setting $\mathcal{R}_{10-d}=0$, we see that for a de Sitter solution we minimally require a RR $F_q$ flux with $q<5$ together with  $H_3$ flux to solve eq. (\ref{constraintsm}). Indeed, using (\ref{constraintsm}) we can rewrite the $d$-dimensional Einstein trace as  
\begin{equation}\label{ddimtraceiso}
    \mathcal{R}_d=\frac{1}{2}(\ab H_3\ab^2_\Int-\ab H_7\ab^2_\Int)\, ,
\end{equation}
which clearly coincides with the dilaton equation (\ref{10ddilateqsm}) once the internal curvature is set to zero.
We can thus state the following no-go: 
\begin{nogo}\label{nogointernal}
No de Sitter solution in any $d$  if $H_3=F_q=0$$\,,\quad q<5$\,.
\end{nogo}
\noindent The minimally required RR flux for a de Sitter solution is then chosen among  $F_0$, $F_2$ and $F_4$  in IIA and $F_1$, $F_3$ in IIB  and such flux must  be accompanied by  $H_3$.
A non-vanishing $H_3$ flux puts a lower-bound on the dimension $n$ of the Scherk-Schwarz torus, $n\ge 3$,  and consequently an upper-bound on the spacetime dimension where de Sitter solutions can be found, $3\le d\le 7$. Hence,  \textbf{dS No-go} \ref{nogointernal} in turn implies:
\begin{nogo}\label{nogod89}
There are no de Sitter solutions in $d=8,9$ spacetime dimensions\footnote{Clearly the trivial case $d=10$ is automatically excluded as it would correspond to the decompactification limit where supersymmetry is recovered and the Casimir term vanishes.}.
\end{nogo}
\noindent Clearly, one can look for de Sitter solutions where other fluxes in addition to those in \textbf{dS No-go} \ref{nogointernal} are present as well. In that case, the latter should constitute the largest contribution to the energy-density for a de Sitter solution to be possible at all. 

In all the other cases, one can only obtain Minkowski and anti de Sitter solutions.
For example, Minkowski solutions could  be minimally supported by the interplay of the 1-loop Scherk-Schwarz term with: $F_5$ in $d=5$,  the pure RR combination $F_4/F_6$ in $d=4$, and by the pure NSNS combination $H_3$/$H_7$ and its RR counterpart $F_3$/$F_7$ in $d=3$. However, considerations on the moduli stabilisation can further exclude some of these configurations.  Anticipating the next section, any RR flux term in the effective potential shares exactly the  same $\phi_d$-dependence as the 1-loop term, $\phi_d$ being the universal modulus whose vev defines the string-coupling of the lower-dimensional Einstein-frame theory. As such, in the presence of a pure RR flux configuration  competing with the 1-loop term, this modulus could not in general be fixed.
Anti de Sitter solutions are possible only in $d=3$ as they minimally require a non-vanishing $H_7$ flux and a RR $F_q$ flux with $q>5$, otherwise (\ref{constraintsm}) forces (\ref{ddimtraceiso}) to be non-negative.

\subsubsection{Product compactification}\label{10secan}
In this case, the internal space, $\mathcal{C}^{10-d}=Y^m\times \mathcal{T}^n_{\text{ss}}$, is a direct product of a  general, compact Euclidean space $Y^m$, with curvature $\mathcal{R}_Y$, and the flat Scherk-Schwarz torus $\mathcal{T}_{\text{ss}}^n$.  The internal curvature is then set exclusively by the $Y$-space, $
    \mathcal{R}_{10-d}=\mathcal{R}_Y$
and from (\ref{rdrint}) a positive (negative) internal curvature will contribute negatively (positively) to the spacetime curvature. In this background,  fluxes $H_3$ and $F_q$ can be decomposed according to (\ref{fluxintdeco}) in terms of (sums of) $H_3^{(p_3)}$, $F_q^{(s_q)}$ flux components with legs solely along $Y^m$ ($p_3=s_q=m$), along $\mathcal{T}^n_{\text{ss}}$ ($p_3=s_q=0$) or in both directions $(p_3=1,2\,,s_q=1,\dots,m-1)$.
To begin with, we  show that for a de Sitter solution  $\mathcal{R}_Y$ has to be positive. Then, in analogy to what has been done in the single-factor compactification case, we will individuate the necessary flux components $H_3^{(p_3)}$, $F_q^{(s_q)}$ for de Sitter solutions.

To this purpose, we first extract the $m$-dimensional and $n$-dimensional Einstein traces from the trace-reversed Einstein equation (\ref{tracerevloop}), 
exploiting the flux decomposition (\ref{fluxdeco}).
When performing the traces, some precise contractions of flux terms  appear for which we exploit the identities listed in Appendix \ref{app:comps}. Using also eq. 
 (\ref{constraintsm}), we eventually obtain the IIA/B $m$-dimensional Einstein traces in the form 
\begin{equation}\label{mcurv}
2\,\mathcal{R}_Y^{\text{IIA/B}}=\sum_{p_3} p_3\ab H_3^{(p_3)}\ab^2_\Int+g_s^2\sum_{q=0}^{10-d}\sum_{s_q}\left(s_q-\frac{m}{2}\right)\ab F_q^{(s_q)}\ab^2_\Int+\frac{2m}{d}\,\kappa^2_{10}g_s^2\,T^{\text{ss}}_{(d)}\,.
\end{equation}
Let us notice that the Scherk-Schwarz term contributes positively to both $\mathcal{R}_d$ and $\mathcal{R}_Y$. 
The trace above can be suggestively rewritten in terms of $\mathcal{R}_d$ using
(\ref{dtraceeqsm}) as
\begin{equation}\label{traceyIIAB}
  \boxed{\mathcal{R}_d=\frac{d}{2m}\left(2\mathcal{R}_{Y}^{\text{IIA,B}}-\sum_{p_3}p_3\ab H_3^{(p_3)}\ab^2_\Int-m\,\ab H_7\ab^2_\Int\delta_{d,3}-g_s^2\sum_{q=1}^{10-d}\sum_{s_q}s_q\ab F_q^{(s_q)}\ab^2_\Int\right)}
\end{equation}
from which we clearly see that -- since the flux terms on the RHS are  (semi-)negative definite -- a de Sitter solution ($\mathcal{R}_d>0$) must have $\mathcal{R}_Y>0$. This can be expressed as a no-go theorem for product compactifications:
\begin{nogo}\label{signry}
    Consider a compactification of the form $\mathcal{M}_d\times Y^m\times \mathcal{T}^{10-d-m}_{\text{ss}}$ with $Y^m$ an m-dimensional Euclidean space and $\mathcal{T}_{\text{ss}}$ the Scherk-Schwarz torus. There are no de Sitter solutions if $\mathcal{R}_Y\le 0$.
\end{nogo}
Let us remark that this result is in  marked contrast with the classical de Sitter landscape where it has been shown e.g in \cite{Andriot:2019wrs, Andriot:2022xjh}, by exploiting the same set of 10d equations, that  de Sitter solutions -- in classical supergravity with localised sources -- necessarily require negative internal curvature.  If a maximally symmetric ansatz for $Y^m$ is further assumed, this amounts to discarding the options $Y^m=\mathcal{H}^m$ and $\mathcal{T}^m$, the $m$-dimensional compact hyperbolic space and torus, respectively, and leaves as a possibility $Y^m=S^m$,  the $m$-dimensional sphere $(m\neq 1)$. The \textbf{dS No-go} \ref{signry}  still applies to the case where the $Y$-space itself is a direct-product of other factors, e.g $Y^m=\prod_i Y^i$, if $\sum_i \mathcal{R}_{Y^i}\le 0$. 
 \vspace{2mm}

We now come to the Einstein equations along the Scherk-Schwarz torus. From the IIA/B traces we have, setting to zero the curvature along the torus, 
\begin{equation}\label{intcurv}
0=\sum_{p_3}(3- p_3)\ab H_3^{(p_3)}\ab^2_\Int+g_s^2\sum_{q=0}^{10-d}\sum_{s_q}\left(q-s_q-\frac{n}{2}\right)\ab F_q^{(s_q)}\ab^2_\Int-\frac{2(10-n)}{d}\,\kappa^2_{10}g_s^2\,T^{\text{ss}}_{(d)}\, .
\end{equation}
As a consistency check, setting $p_3=s_q=m=0$ in the latter and using (\ref{constraintsm}), we match eq. 
 (\ref{intcurviso}), recovering as expected the single-factor compactification case.  
From the above equation, we can express the Casimir term as function of the flux components  $H_3^{(p_3)}$ and $F_q^{(s_q)}$.  Doing so in (\ref{mcurv}), using (\ref{constraintsm}) as well, we obtain
\begin{equation}\label{mcurvpos}
    2\mathcal{R}_Y^{\text{IIA/B}}=\frac{1}{d+m}\left(\sum_{p_3}(d\,p_3+m)\ab H_3^{(p_3)}\ab^2_\Int+2m\ab H_7\ab^2_\Int\delta_{d,3}+d\,g_s^2\sum_{q=0}^{10-d}\sum_{s_q}s_q\ab F_q^{(s_q)}\ab^2_\Int\right)
\end{equation}
and we see that the $Y$-space must actually be positively curved regardless the sign of $\mathcal{R}_d$.  
\vspace{1mm}

By trading the Casimir term for the fluxes in the $d$-dimensional Einstein trace  (\ref{dtraceh7}),  we obtain $\mathcal{R}_d$ exclusively as function of the fluxes; in this way we can read off the sign of the flux component contributions to $\mathcal{R}_d$ and  individuate the minimal set of fluxes for a de Sitter solution, if any. Indeed, using (\ref{intcurv}) together with (\ref{constraintsm}) back in (\ref{dtraceh7}) we eventually obtain
\begin{equation}\label{10ddtracesm1}
\begin{split}
\boxed{\mathcal{R}_{d}=\frac{d}{2(10-n)}\biggl(-(7-n)\ab H_7\ab^2_\Int\delta_{d,3}-\sum_{p_3=0}p_3\ab H_3^{(p_3)}\ab^2_\Int+g_s^2\sum_{q=0}^{10-d}\sum_{s_q}\left(\frac{5-q}{2}-s_q\right)\ab F_q^{(s_q)}\ab^2_\Int\biggr)}
\end{split}
\end{equation}
 As it is evident, the only positive contributions to $\mathcal{R}_d$ come from the RR fluxes, while all the other terms, i.e. the $H_3$ and $H_7$ fluxes, give either a vanishing or a negative contribution. More precisely, we see that only $F_0$, $F_2^{(0)}$, $F_2^{(1)}$ and $F_4^{(0)}$ fluxes in IIA, and $F_1$ and $F_3^{(0)}$ fluxes in IIB, give a strictly positive contribution to $\mathcal{R}_d$. Thus, a de Sitter solution minimally requires one of these fluxes to be non-vanishing.  On the other hand we know that the RR fluxes are related to the $H_7$ and $H_3$ fluxes via the 10d equation  (\ref{constraintsm}). Let us then use the latter in (\ref{10ddtracesm1}), also exploiting the identity (\ref{sqfqid}), to recast $\mathcal{R}_d$ in the equivalent form
\begin{equation}\label{10ddtracesm2}
\boxed{
 \mathcal{R}_{d}=\frac{d}{2(10-n)}\biggl(-(8-n)\ab H_7\ab^2_\Int\delta_{d,3}+\ab H_3^{(0)}\ab^2_\Int-\sum_{p_3\ge 2}(p_3-1)\ab H_3^{(p_3)}\ab^2_\Int-g_{s}^2\sum_{q=0}^{10-d}\sum_{s_q=0}^qs_q\ab F_q^{(s_q)}\ab^2_\Int\biggr)}
\end{equation}
 We now see that all the RR fluxes, and still $H_7$, give a non-positive contribution and that the only  positive contribution,  for which a de Sitter solution would be possible, is given by the $H_3^{(0)}$ component.  This means that $H_3^{(0)}$ must always be present for a de Sitter solution. Combining this observation with \textbf{dS No-go} \ref{signry}, we can hence formulate the generalisations to product compactifications of  \textbf{dS No-go} \ref{nogointernal} and \textbf{dS No-go} \ref{nogod89} for single-factor compactifications:
\begin{nogo}\label{H30}
No de Sitter solution in any $d$ if $H_3^{(0)}=F_q^{(s_q)}=0\,,\quad q<5-2s_q$\,.
\end{nogo}
\begin{nogo} \label{nogodim6}
    No de Sitter solutions from  product compactifications $\mathcal{M}_d\times Y^m\times \mathcal{T}^{10-d-m}_{\text{ss}}$  to $d> 5$ spacetime dimensions.
\end{nogo}
 The different flux components in the product compactifications offer more possibilities to realise Minkowski or anti de Sitter solutions compared to the single-factor compactification case (see discussion below \textbf{dS No-go} \ref{nogod89}). E.g. 
Minkowski solutions could be minimally supported by $H_3^{(0)}$ and $F_3^{(1)}$ as well as $H_3^{(1)}$ and $F_q^{(0)}$ with $q<5$, while anti de Sitter solutions could be supported by $H_3^{(1)}$ and $F_q^{(s_q\ge 1)}$ or $H_3^{(p_3\ge 2)}$ and $F_q^{(s_q\ge 0)}$ for $q<5$. In particular, anti de Sitter solutions could now be realised in $d>3$ spacetime dimensions.
\vspace{2mm}

Before presenting some explicit, fully backreacted solutions, let us make a final general comment regarding the scale separation of such (A)dS solutions using product compactifications with non-trivial curvature. Phenomenologically, one would like to have $\frac{m^2_{KK}}{\ab \Lambda\ab}\gg 1$, where $m_{KK}$ is the KK scale of the compactification and  $\Lambda$ the cosmological constant. It is possible to constrain  this ratio in terms of the  fluxes in the theory \cite{Gautason:2015tig}. For our unwarped compactifications with constant dilaton,  recalling that $\mathcal{R}_Y$ has to be positive (\ref{mcurvpos}) and assuming that $Y$ is approximately isotropic, a good proxy for $m^2_{KK,Y}$ is  given by $\mathcal{R}_Y$; moreover,     we have $\Lambda\sim \mathcal{R}_d$. Using the Einstein equations (\ref{mcurvpos}), and (\ref{10ddtracesm1})   or (\ref{10ddtracesm2}) (and (\ref{constraintsm})) as appropriate,  it is readily checked
that $\frac{\mathcal{R}_Y}{\ab \mathcal{R}_d\ab}\sim \mathcal{O}(1)$, as the same fluxes are appearing both in the numerator and denominator\footnote{As illustration, for AdS solutions minimally supported by $H_3^{(p_3\ge 2)}$ and $F_q^{(s_q)}$ with $q<5$, $s_q\ge 0$ we have
$$\frac{\mathcal{R}_Y}{\ab\mathcal{R}_d\ab}=\frac{1}{d}\frac{(d\, p_3+m)\ab H_3^{(p_3)}\ab^2+2m\ab H_7\ab^2+d\,g_s^2\,s_q\ab F_q^{(s_q)}\ab^2}{(p_3-1)\ab H_3\ab^2+(m+1)\ab H_7\ab^2+g_s^2s_q\ab F_q^{(s_q)}\ab^2}\le \frac{1}{d}\frac{dp_3+m}{p_3-1} \sim \mathcal{O}(1)\,.
$$}. Scale separation of (A)dS solutions is thus  excluded.

\subsection{Some explicit (A)dS solutions} \label{S:concreteegs}
The no-go theorems derived above, using the dilaton eom and traces of the Einsteins equations, can be read as indicating {\it{necessary}} ingredients for  possible de Sitter solutions in the context of type II Scherk-Schwarz single-factor or product compactifications, with a corresponding 1-loop Casimir energy contribution to the action.  We now consider whether these ingredients can actually give rise to de Sitter solutions to the full set of 10d eoms and Bianchi identities by presenting a concrete example, together with  an explicit anti de Sitter solution for comparison.  We will find that, whilst the anti de Sitter solution is self-consistent with the 10d starting assumptions of the $g_s$ and $\alpha'$ expansions, the de Sitter solution presented is not.  In the next section, we will examine more generally self-consistency of any putative de Sitter solution, as well as their stability, by using an effective field theory analysis.

 Before considering our concrete examples, let us make some general observations about our flux backgrounds.  We are interested in the case where the internal NSNS and RR field strengths  contain, or completely reduce to, non-trivial background fluxes, that is the closed but not exact, i.e harmonic, part. Consistency of string theory requires such fluxes to be quantised and the flux quantisation conditions schematically read 
\begin{equation}\label{fluxquantcond}
    \frac{1}{(2\pi\sqrt{\alpha'})^{q-1}}\int_{\mathcal{C}^{q}} F_{q} = n_{q}\in \mathbb{Z}\quad , \quad  \frac{1}{4\pi^2{\alpha'}}\int_{\mathcal{C}^{3}} H_{3} = n_{H_3}\in \mathbb{Z}
\end{equation}
where $\mathcal{C}^q$ is a $q$-cycle of volume $(2\pi\sqrt{\alpha'})^q$ and there is a flux quantisation condition for each $q$-cycle in the internal manifold. Moreover,
in the absence of localized sources and for harmonic fluxes, the Bianchi identities $d F_{q}=H_3\wedge F_{q-2}$ (\ref{fluxbi})  require tadpole-free flux configurations $H_{3}\wedge F_{q-2}=0$. This equation can be respected by distributing appropriately the form legs in the compact space. Since then the fluxes are not constrained by any tadpole cancellation,  the flux numbers $n_{H_3}$, $n_q$ are \textit{unbounded} integers. 

\subsubsection*{$dS_4\times \mathcal{T}_{\text{ss}}^6$} \label{S:dSxT}
As an example of a (4d) de Sitter solution, we consider for simplicity a single-factor\footnote{For product compactifications, there are further constraints on the flux configurations, which must simultaneously satisfy the Bianchi identities and the off-diagonal trace-reversed Einstein equations.  We have not made a systematic search for such configurations because we will see in the following section that -- in any case -- they would not be under parametric control, in contrast to their AdS counterparts.  On the other hand, nor have we have excluded the possibility of
numerical control.} compactification $dS_4\times \mathcal{T}^6_{\text{ss}}$. This solution is supported by $H_3$ and $F_2$ fluxes; therefore it is realised in type IIA. The background is described by the metric
\begin{equation}
    ds^2_{10}=ds^2_4+R^2_{\text{ss}}\,\delta_{ij}\,dy^idy^j
\end{equation}
where $g_{ij}=R_{\text{ss}}^2\delta_{ij}$, with $R_{\text{ss}}$ the radius of the Scherk-Schwarz torus (in units of $\sqrt{\alpha'}$), $\delta_{ij}$  the unit-volume 6d torus metric, and the torus coordinates identified as $y^i\sim y^i+ \ell_s$. The flux ansatz is
\begin{equation}
\begin{split}
    H_3&=\frac{h_3}{2}(dy^1\wedge dy^3\wedge dy^5+dy^1\wedge dy^4\wedge dy^6+dy^2\wedge dy^3\wedge dy^6+dy^2\wedge dy^4\wedge dy^5)\\
    F_2&=\frac{f_{2}}{\sqrt{3}}\left(dy^1\wedge dy^2+dy^3\wedge dy^4+dy^5\wedge dy^6\right) \,, \label{fluxchoice}
\end{split}
\end{equation}
where the flux coefficients are fixed by flux quantisation (\ref{fluxquantcond}) as
\begin{equation}
  h_{3}=\frac{2}{\ell_s}\,n_{H_3}\,,\quad f_{2}=\frac{\sqrt{3}}{\ell_s}\,n_{2}\,.
\end{equation}
It then follows trivially that  $H_3\wedge F_2=0$, hence the $F_4$ Bianchi identity in (\ref{fluxbi}) is satisfied. Moreover, since $H_3$ and $F_2$ are harmonic forms, the remaining flux Bianchi identities in (\ref{fluxbi}) and the flux eoms (\ref{fluxeom}) are automatically satisfied too.  The linear combination of dilaton and 10d Einstein trace equations (\ref{constraintsm})  now fixes 
\begin{equation}\label{dileq}
    \ab H_3\ab^2_\Int=\frac{3}{2}g_s^2\ab F_2\ab^2_\Int\,.
\end{equation} 
The internal Einstein equations are
\begin{equation}
\begin{split}
  0=\frac{1}{4}{H_3}_{ikl}{{H_{3}}_j}^{kl}+\frac{1}{2}g_s^2 \,F_{2ik}{F_{2j}}^k-\frac{g_{ij}}{8}\left(\ab H_3\ab^2_{\Int}+\frac{1}{2}g_s^2\ab F_2\ab^2_\Int-\frac{16}{3}\,\kappa_{10}^2g_s^2\hat{V}\right)\,.
\end{split}    
\end{equation}
From the flux ansatz, it is readily checked  that 
\begin{equation} \label{dSfluxid}
\begin{split}
    {H_3}_{ikl}{{H_3}_{j}}^{kl}&=\ab H_3\ab^2_\Int\,g_{ij}\,,\quad 
    F_{ik}{F_{j}}^k=\frac{1}{3}\ab F_2\ab^2_\Int g_{ij}\,,\\
    \ab H_3\ab^2_\Int&=\frac{h_3^2}{R_{\text{ss}}^6}\,,\quad \ab F_2\ab^2_\Int=\frac{f_2^2}{R_{\text{ss}}^4}
\end{split}    
\end{equation}
and so the internal Einstein equations reduce to their trace
    \begin{equation}\label{intein}
    0=\ab H_3\ab^2_\Int+\frac{5}{6}g_s^2\ab F_2\ab^2_\Int+\frac{16}{3}\kappa_{10}^{2}g_s^2\hat{V} \,.
\end{equation}
The solution to (\ref{dileq}) and (\ref{intein}), consistently also with the dilaton and external Einstein equation, gives
\begin{equation}
    g_s=1.28\, \frac{\ab{n_{H_3}}\ab}{\ab{n_2}\ab^{2/3}}\,,\quad R_{\text{ss}}=0.74\, \frac{1}{{\ab n_2\ab^{1/3}}}\, .
\end{equation}
Clearly, there is no parametric control as a large $n_2$, whilst leading to small string coupling, would unavoidably push the Scherk-Schwarz radius towards sub-Planckian values. We can also see that --for the present example-- numerical control is also excluded; in particular, it is impossible to have control of the $\alpha'$ expansion.   Notice moreover the scaling $g_s\sim n_{H_3}\,R^2_{\text{ss}}$. As we will see in subsection \ref{control}, this scaling turns out to characterise $\textit{any}$ putative de Sitter solution, hence de Sitter with parametric control is definitively excluded. It is also worth noting for the present case of a single-factor compactification that -- even if examples of numerical controlled de Sitter solutions can be found with different flux combinations -- they could be at best only numerically scale separated (c.f. end of subsection \ref{10secan} for the product compactification case).  Indeed, the dilaton equation (\ref{10ddilateqsm}) fixes  ${\mathcal{R}}_4=\frac{1}{2} \ab H_3\ab_{\text{int}}^2$, which together with $\ab H_3\ab_{\text{int}}^2 \sim n_{H_3}^2/(\ell_s^2 R_{\text{ss}}^6)$ implies $(R_{\text{ss}}\,\ell_s)\ell^{-1}_{dS_4}\sim n_{H_3}/R_{\text{ss}}$; this could not be made much less than 1 as necessary for scale separation.

\subsubsection*{$AdS_7\times S^2\times S^1_{\text{ss}}$} \label{S:adSxSxT}
Let us next present an example of an anti de Sitter solution in $d=7$. The metric is given by
\begin{equation}
    ds^2_{10}=ds^2_7+R^2\,\tilde{g}_{ab}\,dy^a dy^b+R^2_{\text{ss}}\,\,dr^2\,,
\end{equation}
where $R$ is the radius of the 2-sphere (in units of $\sqrt{\alpha'}$) with unit-radius metric $\tilde{g}_{ab}$ and $R_{\text{ss}}$ is the radius of the Scherk-Schwarz circle with coordinate $r\sim r+\ell_s$. The solution is again realised in IIA with $H_3^{(2)}$ and $F_2^{(2)}$ fluxes. The flux ansatz is
\begin{equation}
     F_2^{(2)}=f_2 \,{\text{vol}}_2=\frac{1}{2}\,f_2\,\epsilon_{ab}\,dy^a\wedge dy^b \quad \text{and} \quad H_3^{(2)}=h_3\,{\vol}_2\wedge dr\, ,
\end{equation}
with ${\vol}_2$ the 2-sphere unit-radius volume form and 
\begin{equation}
    f_2=\frac{n_2}{\ell_s}\,,\quad h_3=\frac{n_{H_3}}{\ell_s} \, ,
\end{equation}
from the usual flux quantisation conditions. As for the de Sitter solution, the fluxes are harmonic and satisfy $H_3\wedge F_2=0$; therefore the $F_4$ Bianchi identity (\ref{fluxbi}) and all the remaining flux eoms/ Bianchi identities (\ref{fluxeom}) are trivially satisfied too. We are hence left with the dilaton eom and the Einstein equations.  The linear combination of the dilaton eom with the 10d Einstein trace (\ref{constraintsm}) gives again (\ref{dileq}). Due to our flux ansatz, we have
\begin{equation}
\begin{split}
        F_{ac}{F_b}^{c}=\ab F_2\ab^2_\Int\,g_{ab}\, ,&\quad H_{acr}{H_b}^{cr}=\ab H_3\ab^2_\Int g_{ab}\,,\quad   H_{rab}{H_r}^{ab}=2\ab H_3\ab^2_\Int g_{rr}\\
        \ab F_2\ab^2_\Int&=\frac{f_2^2}{R^4}\, ,\quad \ab H_3\ab^2_\Int=\frac{h_3^2}{R^4R_{\text{ss}}^2}\, ,
\end{split}
\end{equation}
and therefore the  internal Einstein equations boil down to the traces along the 2-sphere and the Scherk-Schwarz circle, respectively
\begin{equation}\label{ads7traces}
\begin{split}
      &\frac{3}{8}\ab H_3\ab^2_\Int+\frac{7}{16}g_s^2\ab F_2\ab^2_\Int-\kappa_{10}g_s^2 \hat{V}=\frac{1}{R^2}\, ,\\
      &3\ab H_3\ab^2_\Int-\frac{1}{2}g_s^2\ab F_2\ab^2_\Int+72\kappa_{10}g_s^2 \hat{V}=0\,.
\end{split}
\end{equation} 

The solution to the  10d eoms (\ref{dileq}) and (\ref{ads7traces}) fixes, consistently with the external Einstein equations, the string coupling and compactification radii in terms of the flux numbers $n_{H_3}$ and $n_2$ as, up to numerical coefficients,
\begin{equation}
        g_s\sim n_{H_3}^{5/7}n_2^{-6/7}\,,\quad R_{\text{ss}} \sim n_{H_3}^{2/7} n_2^{-1/7}\,,\quad  R\sim n_{H_3}^{5/7}n_2^{1/7}  \,.
\end{equation}
The solution enjoys parametric control. Indeed, by choosing appropriately the flux numbers $n_{H_3}$ and $n_2$, which recall are unbounded, the solution can lie in the regime of weak coupling and large volume (with $R \gg R_{\text{ss}}$) and thus be self-consistent.  However, separation of scales is trivially excluded, in line with the comments at the end of subsection  \ref{10secan}: the external Einstein equation (\ref{dtraceeqsm}) fixes, up to order 1 numbers, $\mathcal{R}_{7}\sim - \ab H_3\ab^2_\Int$ and  the eoms (\ref{dileq}, \ref{ads7traces}), then imply $R\, \ell_s \,\ell_{AdS_7}^{-1}\, \sim \mathcal{O}(1)$.

\section{\texorpdfstring{$d$}{TEXT}-dimensional EFT analysis, stability and control}\label{ddimEFT}
The 10d discussion carried out so far can be equivalently reproduced in a lower-dimensional framework. In this approach, a de Sitter solution is a critical point $\partial_{\varphi^i}V=0$ with $V>0$, where $V$ is the potential for the scalars $\varphi^i$ of the lower-dimensional effective theory.
This potential arises from the dimensional reduction of the type II 10d supergravity actions (\ref{IIaction}) plus the 1-loop Scherk-Schwarz term (\ref{casimirpot}), on the backgrounds (\ref{isocomp}) and (\ref{anisocomp}) of our interest.  In principle, $V$ would have a rather complicated dependence on all the scalars arising from the reduction,  i.e. the dilaton and all the geometric moduli of the compactification, as well as axions from zero-modes of higher dimensional $p$-forms. Here, however, we follow the method  pioneered in \cite{Hertzberg:2007wc} and focus exclusively on the \emph{universal moduli}. These  are the lower-dimensional dilaton and the one or two (string-frame) internal volume moduli.  Indeed, in section \ref{contruncsec} we will show that studying the behaviour of $V$ along this reduced moduli space, with all the remaining scalars fixed, is enough to  reproduce the 10d no-go theorems of section \ref{10dsol}. We moreover  extend  the no-go theorems to quasi-de Sitter ``slow-roll'' solutions in section \ref{slowroll}.

Having an EFT description is also useful for addressing the question of (perturbative) stability. Indeed, in the lower-dimensional language, stability is related to the sign of the eigenvalues of the Hessian of the potential around a given  critical point. The stability analysis of the classical de Sitter landscape  has been extensively performed  in the slice of moduli space parametrised by the (semi-)universal moduli\footnote{Here, semi-universal moduli refers to both the universal moduli, i.e. the dilaton and overall volume modulus, and the moduli parametrising the volumes of cycles wrapped by the localised sources.}  and it was conjectured \cite{Danielsson:2012et}  that a tachyon always arises among these modes. In the same spirit, in section \ref{pertinstsec} we will perform  this stability analysis for our model and conclude that any de Sitter critical point in $d\ge 4$ spacetime dimensions, if it exists at all, has necessarily a tachyon among the universal moduli. Since, in many instances, these moduli moreover represent a subset of a consistent truncation (see discussion in section \ref{contruncsec}), this amounts to excluding perturbative stability of any de Sitter solution to the full 10d eoms.

For consistency with the string loop expansion and 10d supergravity approximation that have been assumed, valid  solutions must correspond to weak string coupling and large volumes, i.e.
\begin{equation}\label{consistency}
    g_{s}\ll 1\,,\qquad R,R_{\text{ss}}\gg 1\, .
\end{equation}
Note, in particular, that we include the 1-loop Casimir potential but neglect higher-order loop contributions.  The expectation that the 1-loop contribution can be significant despite having $g_s \ll 1$ is thanks to the interplay between the weak coupling and large volume expansions; in particular, we will see that for the product compactifications, the 1-loop contribution is suppressed by $g_s$ and powers of $R_{\text{ss}}$ with respect to the tree-level contributions, but the latter can have a comparable suppression via the length-scale\footnote{If such an effective enhancement of the 1-loop contribution is found, it should then be checked for completeness that higher-loop and higher-derivative corrections are really subleading.} $R \gg R_{\text{ss}}$. 
 Ideally, for maximum control of our approximations, we would like ``parametric control'' of the (vacuum) solutions; that is, the string coupling and the internal volumes can be made,  respectively, arbitrarily small and large by tuning some parameter in the solution, e.g a flux number. Indeed, if  fluxes are not constrained by  the tadpole cancellation conditions, then the associated flux numbers are  \textit{unbounded}: arbitrarily large volumes and small string coupling can be thus achieved by cranking up these fluxes. Whether de Sitter solutions are realisable in this window will be explored in Section \ref{control}. 

\vspace{3mm}

\subsection{The scalar potential}

Let us now derive the EFT and its scalar potential $V$. We decompose the 10d string-frame metric as follows
\begin{equation}\label{10deinsteiniso}
    ds^2_{10}=e^{\frac{4}{d-2}(\phi_d-\braket{\phi_d})}ds^2_{\mathcal{M}_{1,d-1}} +e^{2\chi}d\breve{s}^2_{Y^m}+e^{2\omega}d\breve{s}^2_{\mathcal{T}^{10-d-m}_{ss}}\, ,
\end{equation}
where the second term on the RHS is absent in  single-factor compactifications  (\ref{isocomp}) but present   for  product-compactifications (\ref{anisocomp}). The internal line element is normalised such that
 \begin{equation}
 \begin{split}
\vol_{\text{int}}=\int_Y d^my\sqrt{\breve{g}_Y}\int_{\mathcal{T}_\text{ss}} d^ny \sqrt{\breve{g}_{\mathcal{T}_\text{ss}}}=\ell_s^{10-d}\, ,
 \end{split}
 \end{equation}
 $\omega$ and $\chi$ are the string-frame radions that control the physical internal volumes of the Scherk-Schwarz torus and the $Y$-space, respectively, and $\phi_d$ denotes the $d$-dimensional dilaton 
 \begin{equation}
 \phi_d:= \Phi-\frac{m}{2}\chi-\frac{n}{2}\omega\, .
 \end{equation}
  The Weyl factor $e^{\frac{4}{d-2}(\phi_d-\braket{\phi_d})}$ is needed to find the $d$-dimensional action in the Einstein-frame. The presence of the vev $\braket{\phi_d}$  ensures that  the $d$-dimensional part of Einstein-frame 10d metric (\ref{10deinsteiniso}) coincides with its counterpart in the  string-frame 10d metric (\ref{10ans}) in the vacuum. The vacuum is defined  by the  vevs
\begin{align}
 \label{vevs}
 & e^{\braket{\phi_d}}= \frac{g_s}{R^{m/2}\,R_{\text{ss}}^{n/2}}\,,\quad e^{\braket{\chi}}\equiv R\,,\quad e^{\braket{\omega}}\equiv R_{\text{ss}} \,.
\end{align}
In our conventions, the dimensionless vevs $R$ and $R_{\text{ss}}$ fix the two KK scales of the compactification in string units
\begin{equation}
    \ell_{kk,Y}=R\cdot\ell_s\,,\quad \ell_{kk,T_{ss}}=R_{\text{ss}}\cdot\ell_s\, .
\end{equation}
The moduli $\phi_d$, $\omega$ and $\chi$ are the universal moduli of the  compactification. The single-factor compactification is then recovered by setting $m=0$ and truncating the field $\chi$ in the above and subsequent formulas.

The dimensional reduction of the 10d type II supergravity actions in (\ref{IIaction}) and the 1-loop Scherk-Schwarz runaway potential (\ref{casimirpot}) on the backgrounds (\ref{10deinsteiniso}) generates the $d$-dimensional 1-loop EFT
 action
 \begin{equation}\label{eftaction}
\begin{split}
     S_{d}=\frac{1}{2\kappa^2_d}\int  d^dx\sqrt{-g_d}\biggl(\mathcal{R}_d-K_{ij}\partial_\mu \varphi^i\partial^\mu\varphi^j-V(\varphi^i)\biggr)
     \end{split}
     \end{equation}
 for the scalar fields $\{\varphi^i\}=\{\phi_d,\omega,\chi\}$ minimally coupled to gravity and subjected to a scalar potential $V$. The kinetic terms for $\varphi^i$ arise from the reduction of the 10d dilaton $\Phi$ kinetic term and receive also a contribution from the 10d  Einstein-Hilbert term. In this field basis, the field space metric $K_{ij}$ is diagonal $K_{ij}=\text{diag}(\frac{4}{d-2},n,m)$, and the canonically normalised scalars are then
 \begin{equation}\label{canonfields}
     \hat{\phi}_d=\frac{2}{\sqrt{d-2}}\phi_d\,,\quad \hat{\omega}=\sqrt{n}\,\omega\,,\quad \hat{\chi}=\sqrt{m}\,\chi\, .
 \end{equation}The $d$-dimensional gravitational constant $\kappa^2_{d}$ is given by 
\begin{equation}
    \kappa^2_d:=\frac{\kappa^2_{10}}{\vol_\text{int}}e^{2\braket{\phi_d}}=\frac{\ell_s^{d-2}}{4\pi}e^{2\braket{\phi_d}}\,,
\end{equation}
and defines the Planck mass of the lower-dimensional theory $M_{\text{pl},d}^{d-2}=\kappa^{-2}_{d}$. 
The scalar field vevs (\ref{vevs}), defining the vacuum solution of the quantum theory, correspond to  critical points of the potential $V$.  The  scalar potential $V$ receives two  contributions
\begin{equation}\label{scalarpot}
V=V_{\text{tree}}+2\,\kappa^2_{d}\,V_{\text{1-loop}}\, ,
\end{equation}
and the classical theory is recovered in the $\kappa^2_{d}\rightarrow 0$ limit.

 $V_{\text{1-loop}}$ originates from the Scherk-Schwarz potential $V_{\text{ss}}$ given in (\ref{casimirpot}). In the single-factor compactification, this potential enters already the $d$-dimensional action defined in the 10d string-frame, while in the product compactification a  dimensional reduction on the $Y$-space down to $d$-dimensions is needed. Performing the Weyl transformation to the $d$-dimensional Einstein frame, we obtain a 1-loop contribution to the scalar potential of the $d$-dimensional theory of the form 
\begin{align}
  \label{v1loop}
 &{e^{\frac{4}{d-2}\braket{\phi_d}}}V_{\text{1-loop}}(\phi_d,\omega,\chi)=\frac{\ell_s^m}{e^{2\braket{\phi_d}}} \,e^{\frac{2d}{d-2}\phi_d+m\chi}V_{\text{ss}}(e^\omega)\, .
\end{align}

$V_{\text{tree}}$ incorporates the contributions from the fluxes  and the internal curvature  in the tree-level supergravity.  Dimensional reduction of the 10d NSNS and RR $q$-form kinetic terms and of the 10d Einstein-Hilbert term leads to
\begin{align}
  \label{vtree}
 {e^{\frac{4}{d-2}\braket{\phi_d}}}V^{\text{IIB/A}}_{\text{tree}}(\phi_d,\omega,\chi)=&\,e^{\frac{4}{d-2}\phi_d}\left(\frac{1}{2}\sum_{p_3}\ab H_3^{(p_3)}\ab^2_{\text{int}}-\mathcal{R}_Y\right)+
    \frac{1}{2}e^{\frac{2d}{d-2}\phi_d+m\,\chi+n\,\omega}\,\sum_{q=0}^{10-d}\sum_q\ab F_{q}^{(s_q)}\ab^2_{\text{int}} \nonumber \\
    &+\frac{1}{2}\frac{1}{g_s^4}\,\delta_{d,3}\,\delta_{p_7,m}\,e^{\frac{4(d-1)}{(d-2)}\phi_d+2m\,\chi+2n\,\omega}\ab H_7^{(p_7)}\ab^2_{\text{int}}\, .
\end{align}
Note that the 10d type II actions (\ref{IIaction}) also contain  Chern-Simons terms. However, these terms do not couple to the dilaton, nor to the metric due to their topological nature, hence they do not contribute to $V_{\text{tree}}$ and will be neglected in our discussion.  The curvature term $\mathcal{R}_Y$ is only present in the product-compactification.
The flux conventions are as in section \ref{10dsol}; we dualised external $q$-form fluxes into internal  $(10-q)$-form fluxes and accounting for maximal symmetry, only internal $q\le (10-d)$-form fluxes can then appear in $V_{\text{tree}}$.  Moreover, the $q$-form field strengths  have been further decomposed into sums of components according to (\ref{fluxdeco}), with the integers ($p_3,p_7,s_q$) denoting the number of legs\footnote{\label{p7} We recall that in general, for $q\le 6$, $\{p_3,s_q\}\in\{0,1,\dots,m\}$.
Clearly, since $H_7$ and $F_7$ contributions are only available when $d=3$, we must have $p_7=s_7=m$ in (\ref{vtree}). Moreover, $s_6=m$ and $s_5=m$ is the only available choice for $F_6$ and $F_5$ when $d=4$ and $d=5$ respectively.} that $H_3$, $H_7$ and $F_q$ respectively have along some arbitrary cycles in $Y^m$. The scalar potential contribution from spacetime-filling fluxes needs to be considered carefully; to obtain the correct sign and moduli dependence, we followed the procedure outlined e.g in \cite{Andriot:2022bnb}, \cite{Andriot:2022xjh}. This  is crucial to show that the EFT solution perfectly matches the  10d equations discussed in section \ref{10dsol}.  

The flux terms in (\ref{vtree})  still contain contractions with the internal  metric in (\ref{10deinsteiniso}), hence some  $\omega, \chi$ factors are still implicit. Once pulled out, the full scalar  potential acquires the following explicit moduli  dependence 
\begin{align}
   \label{vflux}
 {e^{\frac{4}{d-2}\braket{\phi_d}}}V(\phi_d,\omega,\chi)=&\,-A_Y\,e^{\frac{4}{d-2}\phi_d-2\chi}+e^{\frac{4}{d-2}\phi_d}{\sum_{p_3}}f_{p_3}^2\,e^{-2 p_3\chi-2(3-p_3)\omega}\nonumber\\
 &+f^2_{p_7}g_{s}^{-4}\,e^{8\phi_d}\delta_{d,3} + e^{\frac{2d}{d-2}\phi_d+m\chi+n\omega}\,\sum_{q=0}^{10-d}{\sum_{s_q=0}^q}f_{s_q}^2\,e^{-2s_q\chi-2(q-s_q)\omega}\nonumber \\
 &-A_{\text{1-loop}}\,e^{\frac{2d}{d-2}\phi_d+m\chi-(10-n)\omega}\, .\, 
\end{align} 
Notice that the coefficients $f_{p_3}$, $f_{s_q}$, $f_{p_7}$, $A_{Y}$ and $A_{\text{1-loop}}$ do not depend on the universal moduli $\phi_d$, $\omega$, $\chi$ and that they are all positive. Indeed,
 the flux contributions to the scalar potential are non-negative, as they arise from sums of squares. For product-compactifications, $A_Y$   depends on the  curvature of the $Y$-space. We know from \textbf{dS No-go} \ref{signry} that a de Sitter solution requires a positively-curved $Y$-space, henceforth  we consider $A_Y>0$. The coefficient of the 1-loop term is also positive and given by $A_{\text{1-loop}}=\frac{3\,\cdot\, 2^{8+2n}}{\pi^6\,n^5}\,\frac{1}{\ell_s^2}$. Moreover, the flux coefficients depend on more specific details of the compactification that we are here neglecting, e.g. the non-universal moduli associated to cycle volumes. In the case of topologically non-trivial background fluxes, they are proportional to the flux numbers associated to string-theory flux quantisation conditions through these cycles. The potential for single-factor compactifications is trivially recovered from (\ref{vflux}) by setting $m=p_3=p_7=s_q=A_{Y}=\chi=0$.

Finally, it is useful to bear in mind that the potential in (\ref{vflux}) has the form of a sum of exponentials, $V = \sum_a  A_a e^{\sum_i\alpha_{a i} \phi_i}$.  For such a potential, it is easy to show\footnote{Indeed, with only $n$ independent terms for $n$ fields, the $n$ equations of motion, $0=\sum_a A_a \alpha_{aj} e^{\sum_i \alpha_{ai}\phi_i}=\sum_a \alpha_{aj} \hat{A}_a$ (where $a=1,\dots, n$, $i, j=1,\dots n$, and we have defined $\hat{A}_a \equiv A_a e^{\sum_i \alpha_{ai}\phi_i}$), can be used successively to eliminate $\hat{A}_1$, $\hat{A}_2$, and so on, until in the final equation $\hat{A}_n$ drops out and we arrive at a constraint on the $n^2$ parameters $\alpha_{ai}$.  Therefore, in the end, there are only $n-1$ constraints on $n$ fields.} that we need at least $n+1$ independent terms ($a=1,\dots,n+1$) to stabilise $n$ fields ($i=1,\dots, n$).

\subsection{Consistent truncations}\label{contruncsec}
A priori, the lower-dimensional EFT just defined may be a consistent truncation to the 10d theory, or a good low-energy approximation, or both, or neither.  For example, as we have pointed out in subsection \ref{S:concreteegs}, compactifications on product manifolds with the $Y$-space sufficiently isotropic so that $m_{KK,Y} \sim 1/(R\ell_s)$ have no separation of scales between the $d$-dimensional spacetime curvature and the KK scale; in this case, the EFT presented is not a good low-energy approximation.  However, we will now show that the eoms for the EFT are equivalent to the 10d dilaton eoms and traced Einstein equations that we considered in section \ref{10dsol} (to derive our de Sitter no-go theorems).  If, moreover, the internal space and fluxes required to source the EFT potential can be chosen in such a way as to solve the 10d flux eoms, Bianchi identities and individual components of the Einstein equations -- as was the case for the both concrete examples in subsection \ref{S:concreteegs} -- then the EFT corresponds to a consistent truncation.  In some cases -- e.g. reductions on group manifolds such as torii and $S^3$ -- the consistency can also be simply argued for using the symmetry properties of the compactification\footnote{Indeed, in such reductions, the left-invariant modes, which note do not necessarily have to be only the massless ones, are singlets under the group action; as such, they do not mix with other modes and hence those other modes can be consistently truncated.}.

The EFT eoms are  the scalar field equations and the Einstein equations. The former stems from the extremisation of the scalar potential, giving three equations for the  three vevs (\ref{vevs})
\begin{equation}
\label{criticalpointeq}
    \frac{\partial}{\partial\varphi^i} V\vert_{\text{crit}}=0\,,\quad   \,\varphi_i=\{\phi_d,\omega,\chi\}\, .
\end{equation}
In detail, they read (with the RHS evaluated at the vevs, i.e on-shell)
\begin{equation}\label{criticalpointan}
    \begin{split}
   (d-2)\partial_{\phi_d}V\vert_\text{crit} &=\, -4\,\mathcal{R}_Y+d\,g_{s}^2\,\sum_{q=0}^{10-d}\ab F_q\ab^2_\Int+2\ab H_3\ab^2_\Int+4\ab H_7\ab^2_\Int\delta_{d,3}+4\,d \,\kappa^2_{d}\,V_{\text{1-loop}}\\
\partial_{\omega}V\vert_\text{crit}&=\,\,\frac{g_{s}^2}{2}\sum_{q=0}^{10-d}{\sum_{s_q}(n-2(q-s_q))\ab F_q^{(s_q)}\ab^2_\Int}-{\sum_{p_3}(3-p_3)\ab H_3^{(p_3)}\ab^2_\Int}-2\,(10-n)\kappa^2_{d}\,V_{\text{1-loop}}\\
   \partial_{\chi}V\vert_\text{crit} &=\, 2\, \mathcal{R}_Y+\,\frac{g_{s}^2}{2}\,\sum_{q=0}^{10-d}{\sum_{s_q}(m-2s_q)\ab F_q^{(s_q)}\ab^2_\Int}-{\sum_{p_3}p_3\,\ab H_3^{(p_3)}\ab^2_\Int}+2\,m \,\kappa^2_{d}\,V_{\text{1-loop}}\,.
    \end{split}
\end{equation}
  The on-shell value of the potential, $V\vert_{\text{crit}}$, sets the value of the $d$-dimensional cosmological constant.
 Indeed, the Einstein equations, which for a maximally symmetric solution reduce to the trace, give  
\begin{equation}\label{lowerein}
    \mathcal{R}_d=\frac{d}{d-2}V\vert_{\text{crit}}\, .
\end{equation}
Overall, the EFT provides us with four equations. On the 10d side,  in section \ref{10dsol}, to derive the no-go theorems we have only dealt explicitly with four of the 10d equations, namely  the 10d dilaton equation of motion $D=0$ in (\ref{10ddilateqsm}), the $d$-dimensional  Einstein trace $E_d=0$ in the form of (\ref{dtraceh7}) and the two internal Einstein traces $E_n=0$ (\ref{intcurv}) and $E_m=0$ (\ref{mcurv}).

 It is then readily checked that the equation $\partial_{\phi_d} V\vert_{\text{crit}}=0$  and the lower-dimensional Einstein equation (\ref{lowerein}) are related to $D$ and $E_d$  as follows 
 \begin{equation}\label{cons}
     4E_d-2D=(d-2)\partial_{\phi_d}V\vert_\text{crit}\,\quad\,-2E_d+\frac{d}{2}D=-d\,V\vert_{\text{crit}}+\mathcal{R}_d(d-2)\, .
 \end{equation}
 We are left with the $\partial_{\chi}V\vert_{\text{crit}}$ and $\partial_{\omega}V\vert_{\text{crit}}$ equations. As might be naively expected, extremising the scalar potential with respect to $\chi$ and $\omega$, which are the EFT scalar fields  controlling the physical sizes of the compactification, is equivalent to solving the 10d internal partial trace equations $E_m=0$ in (\ref{mcurv}) and $E_n=0$ in (\ref{intcurv}); indeed it holds that
 
\begin{equation}\label{cons2}
    E_m=\partial_\chi V\vert_\text{crit}\,,\quad E_{n}=\partial_\omega V\vert_\text{crit}\, .
\end{equation}
Since the dS no-go theorems derived in section \ref{10dsol} involved linear combinations of the 10d equations (\ref{10ddilateqsm}), (\ref{dtraceh7}),  (\ref{mcurv}) and (\ref{intcurv}), then by virtue of (\ref{cons}), (\ref{cons2}) one can  verify that the same dS no-go theorems can be reproduced in the lower-dimensional theory.

\subsection{Perturbative stability}\label{pertinstsec}
Since  supersymmetry is spontaneously broken in our setup, we expect any (vacuum) solution to be unstable at some level; even if perturbative instabilities can be avoided,  non-perturbative decay processes such as bubble nucleation will occur.
Here we show that de Sitter solutions in $d\ge 4$ dimensions are  plagued by instabilities already at the perturbative level.

   Perturbative stability of de Sitter solutions requires the mass spectrum of the fluctuations around the given solution to be free of tachyons. However, obtaining the mass spectrum by expanding the 10d eoms to first order in the fluctuations  is rather cumbersome. Since we are dealing with consistent truncations (see section \ref{contruncsec}), the instability analysis of the 10d solution can be carried out equivalently -- but with much less effort -- in the lower-dimensional theory.  In particular, the reduced mass matrix associated to those modes that have been kept in the consistent truncation will be a diagonal block of the full mass matrix before the truncation. Finding an instability in the set of consistent truncation modes is therefore enough to rule out the stability of the full 10d solution. The vice-versa is not  true as this is a necessary but not sufficient condition for stability; indeed, the lemma\footnote{\label{lemma}\textbf{Lemma} \cite{Andriot:2020wpp}. Let $M$ be a square symmetric matrix of finite size, and $A$ an upper left square block
of $M$. Let $\mu_1$ be the minimal eigenvalue of $M$ and $\alpha$ any eigenvalue of $A$. Then one has
$\mu_1 \le \alpha$. See \cite{Andriot:2020wpp} for details and proof.}  in section 3.3 of  \cite{Andriot:2020wpp} shows that adding more fields lowers the Hessian minimal eigenvalue and thus takes solutions towards instability.
   
   In the language of the lower-dimensional theory, the problem then translates into determining whether the Hessian of the scalar potential around a given critical point admits a negative eigenvalue. This problem is further simplified by advocating the \emph{Sylvester's criterion}, according to which an $n\times n$ hermitian matrix is positive-definite if all its principal minors have positive determinant. 
   For classical de Sitter solutions with localised sources, it has been conjectured in \cite{Danielsson:2012et} that a tachyon always arises in the space spanned by the lower-dimensional dilaton and moduli associated to diagonal components of the metric. We will now show using our EFT, with 1-loop Casimir contribution, that there is a universal tachyon for de Sitter solutions in $d\ge 4$ dimensions, which are therefore unstable. 

    Using  the field space metric $K_{ij}$ and the scalar potential $V$ defined in (\ref{eftaction}), we define the  mass-matrix
\begin{equation}\label{massmatrix}    
{M^i}_j=K^{ik}\,\nabla_k\,\partial_j V\vert_{\text{crit}}\,,
\end{equation}
where the indices run over the  fields $\{\phi_d,\omega\}$ in the single-factor compactification and $\{\phi_d,\omega,\chi\}$ in the product compactification. The mass-squareds are the eigenvalues of ${M^i}_j$ evaluated at the critical point defined by  (\ref{criticalpointeq}). Since the field space metric $K^{ij}$ is positive-definite, positivity of ${M^i}_{j}$ requires positivity of $\partial_i\partial_j V\vert_{\text{crit}}$. For the sake of analytical control,  we rely on \textit{Sylvester's criterion} and start by analyzing the smallest principal minors, i.e the diagonal entries.  A negative diagonal entry is then sufficient to conclude that ${M^i}_j$ has a negative eigenvalue. 

Let us analyse for instance  $\partial^2_{\phi_d}V\vert_{\text{crit}}$:
\begin{equation}
    (d-2)^2\partial^2_{\phi_d}V\vert_{\text{crit}}= 4\,d^2\kappa^2_{d}\,V_{\text{1-loop}}-4\mathcal{R}_Y+2\ab H_3\ab^2_\Int+d^2\,\frac{g_s^2}{2}\sum_q \ab F_q\ab^2_\Int+8\ab H_7\ab^2_\Int\, .
\end{equation}
We can use the $\partial_i V\vert_{\text{crit}}=0$ equations (\ref{criticalpointan}) to eliminate the curvature and 1-loop terms at the critical point; then, using (\ref{10ddtracesm2}) and (\ref{lowerein}), we obtain eventually
\begin{equation}
\begin{split}\label{par2phi}
       (d-2)^2 \partial^2_{\phi_d}V\vert_{\text{crit}}
       {=}\ab H_7\ab^2_\Int\delta_{d,3}-2d\,V\vert_{\text{crit}} \,.
\end{split}
\end{equation}
For a de Sitter solution $V\vert_{\text{crit}}>0$ and the right hand side is negative-definite for $d\ge 4$, as well as for $d=3$ without $H_7$ flux. Therefore, by Sylvester's cryterion
this is enough to conclude that
\begin{equation}
    \textit{Any de Sitter  solution in $d\ge 4$ and in $d=3$ if $H_7=0$ is perturbatively unstable.}
\end{equation}
 Note that (\ref{par2phi}) points out the presence of a \emph{universal tachyon} in the spectrum of any de Sitter solution in $d\ge 4$ that corresponds to a consistent truncation, i.e  an instability that does not depend on the details of a particular solution e.g the choice of fluxes and/or curvature, but  rather  an intrinsic feature of this class of (quantum) de Sitter solutions.  
    
 In $d=3$, stability of de Sitter solutions cannot be excluded directly from (\ref{par2phi}) if $H_7$ flux is switched on. In this case, to derive sufficient conditions for instability one has to proceed further on with Sylvester's cryterion. For the $d=3$ \emph{single-factor  compactification}, the mass-matrix (\ref{massmatrix}) $M_{\phi_d,\omega}$ is two-dimensional and the second diagonal entry $\partial^2_\omega V\vert_{\text{crit}}$ can easily  be shown to be positive if $V\vert_{\text{crit}}>0$ (see below and eq. (\ref{d2omegaV})).
One then has to inspect the determinant of the mass matrix $M_{\phi_d,\omega}$ itself. Indeed,  since $\text{Tr}M_{\phi_d,\omega}>0$ holds in this case, then one has that $\lambda_-<0 \iff  \text{det}M_{\phi_d,\omega}<0$ \footnote{ For a $2\times2$ matrix $M$ the smallest eigenvalue is given by $\lambda_-=\frac{1}{2}\left(\text{Tr}(M)-\sqrt{(\text{Tr}M)^2-4\text{det}M}\right)\,$.}.
 The determinant is easily found to be 
 \begin{equation}    
\text{det}M_{{\phi_d},\omega}=-144\,V\vert_{\text{crit}}\left(3\,V\vert_{\text{crit}}+\ab H_7\ab^2_\Int\right)+2g_s^2\sum_q(5-q)^2\ab F_q\ab^2_\Int\left(\ab H_7\ab^2_\Int-6\,V\vert_{\text{crit}}\right)\,,
\end{equation}
from which we derive the sufficient condition for perturbative instability of $d=3$ de Sitter solutions:
\begin{equation}
    V\vert_{\text{crit}}>\frac{1}{6}\ab H_7\ab^2_\Int \,.
    \end{equation}
    In other words, it may be possible to find a metastable de Sitter solution in $d=3$ if a configuration can be found with $ V\vert_{\text{crit}}<\frac{1}{6}\ab H_7\ab^2_\Int$.
For the \emph{product compactification}, we have for the diagonal entries (the single-factor case can be recovered by taking $m=0$,  $H_3^{(p_3)}=0$ for $p_3\neq 0$, and $F_q^{(s_q)}=0$ for $s_q \neq 0$) 
\begin{equation}\label{d2omegaV}
    \begin{split}
\partial^2_{\omega}V\vert_{\text{ext}}=&\frac{2}{d-2}(m+d)^2\,V_\text{crit}-\ab H_7\ab^2_\Int(m(6+11m)-9)\delta_{d,3}+2g_s^2\sum_q\sum_{s_q}(5-q+s_q)^2\ab F_q^{(s_q)}\ab^2_\Int\\
&-6\ab H_3^{(0)}\ab_\Int^2(d+m-3)-4\ab H_3^{(1)}\ab^2_\Int(d+m-2)-2\ab H_3^{(2)}\ab^2_\Int(d+m-1)
\end{split}
\end{equation}
and
\begin{equation}\label{d2chiV}
\begin{split}
    \partial^2_\chi V\vert_{\text{crit}}=&-\frac{2m(m+2)}{d-2}V\vert_{\text{crit}}-m(m+2)\ab H_7\ab^2_\Int\delta_{d,3}+2\sum_{p_3=0}p_3(p_3-1)\ab H_3^{(p_3)}\ab^2_\Int\\
    &-\frac{g_s^2}{2}\sum_q\sum_{s_q}(m(m-1)+6\,s_q)\ab F_q^{(s_q)}\ab^2_\Int\,.
\end{split}    
\end{equation}
We see that the diagonal entries of the mass matrix are no longer (semi-)positive definite, making more room for instabilities; this is to be expected given the lemma in section 3.3 of \cite{Andriot:2020wpp} (see footnote \ref{lemma}) and the fact that we have introduced an additional field with respect to the single-factor compactification. Note in particular that ($d=3$) minimal de Sitter solutions, i.e those minimally supported by $H_3^{(0)}$, have $\partial^2_\chi V<0$ and therefore they are tachyonic. 

In contrast to $d\ge 4$ dS solutions, it is not possible to exclude a priori the stability of AdS solutions. Rather, the stability analysis must be carried out case by case using the data of each particular solution. For example, for the concrete $AdS_7$ solution  we presented in (\ref{S:concreteegs}), we have from (\ref{d2omegaV}) and using the eoms  (\ref{dileq}, \ref{ads7traces}) $\partial^2_{\hat{\omega}}V=-5.7\ab V_{\text{crit}}\ab <-\frac{6}{5}\ab V_{\text{crit}}\ab \equiv m^2_{\text{BF}}$, hence  the solution turns out to be tachyonic and perturbatively unstable.

\subsection{Weak coupling and large volume expansions}\label{control}
In Section \ref{S:concreteegs}, we identified some explicit de Sitter and anti de Sitter solutions to the 10d eoms; however, we found that the de Sitter example we examined was not parametrically consistent with the weak coupling and large volume expansions that we assumed. We will now show that this is not simply a feature of a particular solution but rather a feature of \textit{any} putative de Sitter solution.  To address whether (anti) de Sitter solutions in the parametrically weakly coupled and large volume regimes are at all possible, we analyse  the structure of the scalar potential. In the language of the lower-dimensional EFT, a parametrically controlled  solution corresponds to  $\nabla V\vert_{\text{crit}}=0$ with the vevs (\ref{vevs}), defined by the critical point of the potential, such that (\ref{consistency}) holds parametrically, e.g in terms of an arbitrarily tunable flux number.  This means that the tree-level potential $V_{\text{tree}}$, from fluxes and curvature, must be made small such that  the 1-loop Scherk-Schwarz potential $V_{\text{1-loop}}$ can compete  to allow a critical point, e.g. by tuning the fluxes, but without leaving the parametric weak coupling and large volumes regime. However, as we will now show, the necessary ingredients for de Sitter solutions individated in Section \ref{10dsol} cannot give parametrically comparable contributions to the EFT potential in the weak coupling, large volume regime; hence, de Sitter solutions cannot lie in the parametrically controlled regime.

\subsubsection{Controlled de Sitter solutions?}

For concreteness, we stick to 4d de Sitter spacetime in this discussion and hence we set $d=4$. Our conclusions apply to the other dimensions as well.

\subsubsection*{Single-factor compactification}
 In this scenario, parametric control is straightforwardly excluded. The general 4d scalar potential (\ref{vflux})  in terms of  the vevs $g_s$ and $R_{\text{ss}}$ as defined in (\ref{vevs}) takes the form 
\begin{equation}\label{potiso}
V(g_s,R_{\text{ss}})=\frac{f^2_{H_3}}{R_{\text{ss}}^6}+g_s^2\sum_{q=0}^{6}\frac{f^2_q}{R_{\text{ss}}^{2q}}-A_{\text{1-loop}}\frac{g_s^2}{R_{\text{ss}}^{10}}\, ,
\end{equation}
with the three contributions given respectively by the NSNS $H_3$ flux, the RR $F_q$ fluxes and the 1-loop Scherk-Schwarz term. Let us note that, being $f^2_{H_3},f^2_q, A_{\text{1-loop}}>0$, the 1-loop term is the only negative term in the potential and thus mandatory to avoid a runaway.

We are interested in the case of non-trivial background fluxes for which the flux quantisation conditions (\ref{fluxquantcond}) fix
\begin{equation}
    f_{H_3}\sim\frac{1}{\sqrt{2}\ell_s}n_{H_3}\,,\quad f_{q}\sim\frac{1}{\sqrt{2}\ell_s}n_{q}\qquad n_{H_3},n_q\in\mathbb{Z}\, .
\end{equation}
Recall, moreover, that we consider tadpole-free flux configurations i.e. the fluxes  are not constrained by tadpole cancellation, so the flux numbers  $n_{H_3}$, $n_{q}$ are unbounded.  This is of utmost importance when, as in the present case, a vacuum solution is attempted by comparing terms at different orders in $g_s$ (string coupling) and $R_{\text{ss}}$ (volume) expansion without generating a Dine-Seiberg runaway.  Note that, in contrast, the other numerical coefficients cannot help in the balance of terms;  in particular, the 1-loop term comes with a fixed coefficient,  $A_{\text{1-loop}}=0.42\,\ell_s^{-2}$ for $n=6$, that cannot be tuned arbitrarily large to  balance the suppression in the weak coupling/large volume limit.   

 In the most general case, obtaining closed form expressions for the vevs $g_s$, $R$ and $R_{\text{ss}}$ from the extremisation of $V$ becomes very hard, if not impossible. And yet, some considerations can still be drawn by considering the scaling behavior of some relevant terms.
By virtue of \textbf{dS No-gos} \ref{signcas} and \ref{nogointernal}, de Sitter solutions minimally require the $H_3$ flux together with a RR $F_q$ flux with $q<5$ and the negative 1-loop Scherk-Schwarz term. Let us then assume only one such RR flux, namely $F_2$ or $F_4$ in IIA\footnote{In the presence of a topologically non-trivial (harmonic) $H_3$ flux and without $O6$-planes, the Roman mass $F_0$ must be set to zero for consistency with the $F_2$ Bianchi identity.} and $F_1$ or $F_3$ in IIB (thus three terms to stabilise two moduli). For a de Sitter solution, these three terms must be of same order. Requiring the $H_3$ flux term to be comparable with the $F_q$ term, and imposing the flux quantisations (\ref{fluxquantcond}), fixes the relation
\begin{equation}\label{h3vsfq}
    g_s^2\sim\frac{n_{H_3}^2}{n^2_q}\,R_{\text{ss}}^{2(q-3)}\,,
\end{equation}
from which we clearly see that unbounded flux numbers are necessary for parametric control on the solution, especially for $q\ge 3$ when -- assuming a very large volume regime --  a parametrically small string coupling could still be reached by cranking up the $n_q$ flux number to compensate for the large volume enhancement.
However, it is straightforward to realize that in the parametric large volume and small coupling limit, the Scherk-Schwarz term cannot ever be at leading order. Indeed, for it to be same order as the  $F_q$ flux term we need
\begin{equation}\label{rssnq}
R_{\text{ss}}\sim n_q^\frac{1}{q-5}   
\end{equation}
and for $q<5$ a parametric large volume limit is obstructed by the flux quantisation conditions (\ref{fluxquantcond}). Lastly, comparing the Scherk-Schwarz term with the $H_3$ term we see that $g_s$ grows with the volume $R_{\text{ss}}$ as 
\begin{equation}\label{gsrss}
    g_s\sim n_{H_3}R_{\text{ss}}^{2}\, .
\end{equation}
Accounting once again for flux quantisation conditions, we clearly see that this excludes parametrically small string coupling at large volume  and thus rules out parametrically controlled de Sitter solutions.

\subsubsection*{Product compactification}
The situation changes if we consider the product compactifications. In this case, we need the  third scalar $\chi$, which parametrises (the fluctuations of) the volume of the $Y$-space\, and the  $\alpha$'  perturbative expansion is now governed  by two vevs,  $R$ and $R_{\text{ss}}$ (see (\ref{vevs})). It will be useful to introduce \textit{anisotropy coefficient} $\delta$  to parametrise the internal hierarchy
  \begin{equation}
      \delta:=\frac{R_{\text{ss}}}{R}\,.
  \end{equation}
  We can then rewrite the 4d scalar potential (\ref{vflux}) for the three vevs $g_s$, $R$, $R_{\text{ss}}$ in terms of $g_s$, $R_{\text{ss}}$ and $\delta$  
\begin{equation}\label{4terms}
\begin{split}
 V(g_s,R_{\text{ss}},\delta)=&-A_Y\frac{\delta^2}{R_{\text{ss}}^2}+\sum_{p_3}n_{H_3,p_3}^2\frac{\delta^{2p_3}}{R_{\text{ss}}^{6}}+g_s^2\sum_{q=0}^{10-d}\sum_{s_q}\,n_{q,{s_q}}^2\frac{\delta^{2s_q}}{R_{\text{ss}}^{2q}}
    -A_{\text{1-loop}}\,\frac{g_{s}^2}{R_{\text{ss}}^{10}}\,.
\end{split}    
\end{equation}
The first term accounts for the  positive curvature $(A_Y>0)$ of the $Y$-space as dictated by \textbf{dS No-go} \ref{signry}. Even though the Scherk-Schwarz term is no longer the only negative term in the potential, the requirement of it being leading  at parametric large volumes and weak coupling still holds;  if it were instead subleading the no-go theorem (\ref{dtraceh7}) would forbid any parametrically controlled de Sitter solution.  As such, requiring the curvature term to be of the same order as the Scherk-Schwarz term, that is 
\begin{equation}
    R_{\text{ss}}\sim g_s^{1/4}\,{\delta^{-1/8}}\, ,
\end{equation}
in the weak coupling and large volume limit requires 
\begin{equation}\label{comp} \delta\ll 1 \,.
\end{equation}
  A \textit{large anisotropy in the internal space} is therefore a necessary (not yet sufficient) ingredient to obtain large volume and weakly coupled solutions (\ref{consistency}) out of the 1-loop effect, thus circumventing the Dine-Seiberg problem. Indeed, even though the 1-loop term is suppressed both at weak coupling and in the large $R_{\text{ss}}$ limit, the tree-level curvature term is crucially  suppressed  in a different, large $R$ limit such that we can in principle exploit this latter suppression to make the two  terms of the same order.  This consideration holds regardless the detail of the NSNS and RR flux contributions, since it comes from the comparison between the curvature and the one-loop terms that must always be present.  
  
  A small $\delta$ (\ref{comp}) also helps in balancing  other terms in the potential. Indeed, the relation in (\ref{rssnq}) from balancing the $F_q$ flux term against the 1-loop Casimir now generalises to
\begin{equation}
    R_{\text{ss}}\sim (n_{q,s_q}\,\delta ^{s_q})^{\frac{1}{q-5}}\, .
\end{equation}
A parametrically large $R_{\text{ss}}$ thus becomes possible for $q<5$, $s_q\neq 0$,  and a  small $\delta$. Similarly, eq. (\ref{gsrss}) from balancing the $H_3$ flux term against the 1-loop Casimir becomes
\begin{equation} \label{H3vs1loop}
    g_s\sim n_{H_3,p_3}\, R_{\text{ss}}^2\,\delta^{p_3}\, ,
\end{equation}
 and a parametrically small coupling can be obtained by  overcoming the $R_{\text{ss}}$ enhancement with the $\delta$ suppression, when $p_3\neq 0$.

So far, so good, but now we must recall  the \textbf{dS No-go} 
 \ref{H30}, which tells us that de Sitter solutions minimally require a non-vanishing $H_3^{(0)}$ flux component contribution. This means that the $p_3=0$ flux term in (\ref{4terms})  should be at leading-order together with the one-loop term. However, we can see from from (\ref{H3vs1loop})  that this cannot happen simultaneously at parametrically large volume and weak coupling, because the $p_3=0$ flux term lacks  the $\delta$ suppression. Indeed, this brings us back to the relation (\ref{gsrss}),
which shows that the parametric weak coupling limit implies sub-Planckian $R_{\text{ss}}$.
This conclusion also applies to the more general case where other $H_3^{(p_3\ge 1)}$  flux contributions can be present; by (\ref{10ddtracesm2}) de Sitter solutions simply require these other flux terms to be subdominant to the $H_{3}^{(0)}$ contribution.  In this way, we have shown that it is impossible to find de Sitter solutions that are parametrically and simultaneously at weak coupling and large volume.

The same conclusion on de Sitter solutions can be reached exploiting the 10d equations of motion directly. Let us start from the expression for $\mathcal{R}_d$ in (\ref{dtraceh7}). For a 4d de Sitter solution we must have
\begin{equation}
    4\kappa^2_{10}g_s^2\ab \hat{V}_{\text{ss}}\ab>g_s^2\sum_{q=0}^{6}\sum_{s_q}\ab F_q^{(s_q)}\ab^2_\Int\, .
\end{equation}
 Combining the latter with the 10d equation (\ref{constraintsm}) 
we obtain the following chain
\begin{equation}
    20\kappa_{10}^2g_s^2\ab \hat V_{\text{ss}}\ab > 5 g_s^2\sum_{q=0}^6\sum_{s_q}\ab F_q^{(s_q)}\ab^2_\Int>2\ab H_3\ab^2_\Int\ge 2 \ab H_3^{(0)}\ab^2_\Int\,,
\end{equation}
where in the last part we further used that for de Sitter solutions the $H_3$ flux contribution should  mainly come from the $H_3^{(0)}$ term (see \textbf{dS No-go} \ref{H30}). 
The scaling in (\ref{gsrss}) then follows by considering the extrema of the chain and the flux quantisation conditions (\ref{fluxquantcond}).

\subsubsection{Parametrically controlled anti de Sitter solutions}
Although parametrically controlled de Sitter solutions are impossible in our set up, parametrically controlled non-supersymmetric anti de Sitter solutions are  straightforward.    We have seen that  parametric control needs the leading-order flux terms to come with an $\delta$-suppression factor. Such terms are e.g $H_3^{(p_3\ge 2)}$ and $F_{q}^{(s_q\ge 2)}$,  which can source anti de Sitter solutions. Indeed we can find some parametrically controlled large volume and weakly coupled AdS solutions by considering e.g IIA $H_3^{(2)}$ and $F_{2}^{(2)}$ fluxes in
(\ref{4terms}), as for the concrete AdS solution presented in subsection \ref{S:adSxSxT}. The solution, in terms of the unbounded flux numbers $n_{3}$ and $n_{2}$, is given by

\begin{equation}
        g_s\sim n_{H_3}^{5/7}n_2^{-6/7}\,,\quad R_{\text{ss}} \sim n_{H_3}^{2/7} n_2^{-1/7}\,,\quad  R\sim n_{H_3}^{5/7}n_2^{1/7}  \,.
\end{equation}
for which $g_s \ll 1$ and $R_{\text{ss}}\gg R \gg 1$ can easily be achieved, thus circumventing the Dine-Seiberg problem.

\subsection{No-go theorems for slow-roll acceleration}\label{slowroll}
We have shown that the ingredients of curvature, fluxes and a Casimir potential cannot give rise to parametrically controlled de Sitter solutions; we might therefore expect that nor can they give rise to ``quasi-de Sitter solutions'' with slowly rolling moduli, i.e $V>0$ with a slope satisfying $\epsilon_V \equiv\frac{d-2}{4}\left(\frac{\nabla V}{V}\right)^2 \ll 1$. In fact, the EFT potential allows one to extend the de Sitter no-go theorems derived in section \ref{10dsol} to no-go theorems for slow-roll acceleration {\it{anywhere in the moduli space}} (assuming parametric or numerical control).  Following  the strategy of \cite{Hertzberg:2007wc} (see also
\cite{Andriot:2019wrs, Andriot:2022xjh} for more recent works), 
 we look for inequalities of the type 
\begin{equation} \label{quasinogo}
    a V+\sum_i \hat{b}_i\partial_{\hat{\varphi}^i}V\le 0\,,
\end{equation}
with $\hat{\varphi}^i$ the canonically normalised fields (\ref{canonfields}), $a>0$ and $\hat{b}_i\neq 0$ some 
chosen coefficients. The above leads\footnote{For a proof see Section 3.1 of \cite{Andriot:2019wrs}.} to  a lower bound for the slope $\ab\nabla V\ab$:
\begin{equation}
    \frac{\ab \nabla V\ab}{V}\ge c\,,\quad c=\frac{a^2}{\sum_i \hat{b}_i^2}\,,
\end{equation}
which, if $c\gtrsim\mathcal{O}(1)$, rules out slow-roll.  Note also that when $\partial_{\hat{\varphi}^i}V=0$ a de Sitter no-go theorem is recovered. Starting from our effective potential (\ref{scalarpot}-\ref{vflux}), one can show that the following inequality holds (from now on we absorb the factor $e^{\frac{4}{d-2}\braket{\phi_d}}$ inside $V$)
\begin{equation}
\begin{split}
    4V-(d-2)\partial_{{\phi}_d}V&=(d-2)\left(-e^{\frac{2d}{d-2}\phi_d+m\chi+n\omega}\sum_{q}\ab F_q\ab^2_\Int-2\ab H_7\ab^2_\Int\delta_{d,3}-\frac{4\kappa^2_{d}}{e^{2\braket{\phi_d}}}V_{\text{1loop}}\right)\le 0\\ &\text{if}\quad V_{\text{1-loop}}\ge 0\,,
\end{split}    
\end{equation}
which reproduces \textbf{dS No-go} \ref{signcas}. In terms of the canonically normalised fields,  it gives the $c$-value
\begin{equation} \label{cdS}
    4V-2\sqrt{d-2}\,\partial_{\hat{\phi}_d}V\le 0 \implies
    c=\frac{2}{\sqrt{d-2}}\, .
\end{equation}
For single-factor compactifications, the linear combinations
\begin{equation}
    \frac{2d}{d-2}V-\partial_{{\phi}_d}V=\frac{1}{2}e^{\frac{4}{d-2}\phi_d}\left(\ab H_3\ab^2_\Int-\frac{1}{g_s^4}e^{\frac{4}{d-2}\phi_d+2n\,\omega}\ab H_7\ab^2_\Int\delta_{d,3}\right)\le 0\,\quad \text{if}\quad H_3=0\,,
\end{equation}
\begin{equation}
    \frac{4d}{d-2}\,V+\frac{(d-6)}{2}\partial_{{\phi}_d}V+\partial_\omega V=e^{\frac{2d}{d-2}\phi_d+m\chi+n\omega}\sum_q\ab F_q\ab^2_\Int(5-q)\le 0\,\quad \text{if}\quad F_q=0\,,q<5
    \end{equation}
reproduce the $\textbf{dS No-go}$ \ref{nogointernal} and they both give the same $c$-value:
\begin{equation}
\begin{split}
\frac{2d}{d-2}V-\frac{2}{\sqrt{d-2}}\,\partial_{\hat{\phi}_d}V\le 0\,,&\quad\text{and}\quad \frac{4d}{d-2}V+\frac{d-6}{\sqrt{d-2}}\,\partial_{\hat{\phi}_d}V+\sqrt{10-d}\,\partial_{\hat{\omega}}V\le 0\\
   &\implies c=\frac{d}{\sqrt{d-2}} \,.
\end{split}    
\end{equation}
For product compactifications, the linear combinations
\begin{equation}
\begin{split}
 &V-\frac{1}{8}\frac{d-2}{d+m}(d-2(3+m))\partial_{\phi_d}V+\frac{3}{4}\frac{d-2}{d+m}\partial_\chi V+\frac{1}{4}\frac{d-2}{d+m}\partial_\omega V=  \\
 &=\frac{d-2}{2(10-n)}\biggl(-(7-n)\frac{1}{g_s^4}e^{8\phi_d+2n\,\omega+2m\,\chi}\ab H_7\ab^2_\Int\delta_{d,3}-e^{\frac{4}{d-2}}\sum_{p_3}p_3\ab H_3^{(p_3)}\ab^2_\Int\\
 &\qquad \qquad +e^{\frac{2d}{d-2}\phi_d+n\chi+m\omega}\sum_q\sum_{s_q}\left(\frac{5-q}{2}-s_q\right)\ab F_q^{(s_q)}\ab^2_\Int\biggr)\le 0\,\quad \text{if}\quad F_q^{(s_q)}=0\,,q<5-2s_q
\end{split} 
\end{equation}
and
\begin{equation}
    \begin{split}
        &V-\frac{1}{4}\frac{d-2}{d+m}(m+2)\partial_{\phi_d}V+\frac{1}{2}\frac{d-2}{d+m}\partial_\chi V=\\
        &\qquad \qquad= \frac{d-2}{2(10-n)}\biggl(-(8-n)\frac{1}{g_s^4}e^{8\phi_d+2n\,\omega+2m\,\chi}\ab H_7\ab^2_\Int\delta_{d,3}+e^{\frac{4}{d-2}\phi_d}\ab H_3^{(0)}\ab^2_\Int\\
        &\qquad\qquad-e^{\frac{4}{d-2}\phi_d}\sum_{p_3\ge 2}(p_3-1)\ab H_3^{(p_3)}\ab^2_\Int -e^{\frac{2d}{d-2}\phi_d+n\chi+m\omega}\sum_q\sum_{s_q}s_q\,\ab F_q^{(s_q)}\ab^2_\Int\biggr)\le 0\,\quad \text{if}\quad H_3^{(0)}=0
    \end{split}
\end{equation}
reproduce the \textbf{dS No-go \ref{H30}}. The associated $c$-value is
\begin{equation}
\begin{split}
    &V-\frac{1}{4}\frac{\sqrt{d-2}}{d+m}(d-2(3+m))\partial_{\hat{\phi}_d}V+\frac{3}{4}\frac{d-2}{d+m}\sqrt{m}\partial_{\hat{\chi}}V+\frac{1}{4}\frac{d-2}{d+m}\sqrt{n}\partial_{\hat{\omega}}V\le0\,\quad  \text{and}\\
      & V-\frac{1}{2}\frac{\sqrt{d-2}}{d+m}(m+2)\partial_{\hat{\phi}_d}V+\frac{1}{2}\frac{d-2}{d+m}\sqrt{m}\partial_{\hat{\chi}}V\le0\\[2ex]
      &\qquad\qquad\qquad \implies c=\frac{2(d+m)}{\sqrt{(d-2)(4+m(d+m+2))}}\, .
 \end{split}   
\end{equation}
Lastly, from the linear combination
\begin{equation}
\begin{split}
    \frac{2m}{d-2}V-\frac{m}{2}\partial_{{\phi}_d}V+\partial_{{\chi}}&V=\biggl(-\frac{m}{g_s^4}e^{8\phi_d+2n\,\omega+2m\,\chi}\ab H_7\ab^2_\Int\delta_{d,3}-e^{\frac{4}{d-2}\phi_d}\sum_{p_3}p_3\ab H_3^{(p_3)}\ab^{2}_\Int\\
    &-e^{\frac{2d}{d-2}\phi_d+n\,\chi+m\omega}\sum_q\sum_{s_q}s_q\ab F_q^{(s_q)}\ab^2_\Int+e^{\frac{4}{d-2}\phi_d}\mathcal{R}_Y\biggr)\le 0\,\quad \text{if}\quad \mathcal{R}_Y\le 0
\end{split}
\end{equation}
we recover the \textbf{dS No-go \ref{signry}} and the corresponding $c$-value reads
\begin{equation} \label{csmall}
     \frac{2m}{d-2}V-\frac{m}{\sqrt{d-2}}\partial_{{\hat{\phi}}_d}V+\sqrt{m}\partial_{{\hat{\chi}}}V\le 0\implies c=2\sqrt{\frac{m}{(d-2)(d+m-2)}}
\end{equation}
The strongest bound on the slow-roll parameter $\epsilon_V$ comes from the maximum value of  $c \sim \mathcal{O}(1)$ found above; hence slow-roll acceleration is impossible.

\subsection{Swampland conjectures}

It is interesting to compare the values of $c$ derived above  -- from explicit supersymmetry-breaking string compactifications and valid at (not necessarily asymptotically) weak coupling and large volume -- with the various swampland conjectures.  E.g. the Transplankian Censorship Conjecture argues that $|\nabla V|/V \geq 2/\sqrt{d-2}$ in the asymptotics of field space \cite{Bedroya:2019snp}; requiring that the de Sitter conjecture \cite{Ooguri:2018wrx} be preserved under dimensional reduction also leads to the same bound \cite{Rudelius:2021oaz}.  This bound is verified by the value of $c$ found above in (\ref{cdS}); stronger (and weaker) bounds are also found.

We can also use our concrete string construction to test other swampland conjectures.  In  (\ref{par2phi}), we found a contribution to the lower-dimensional theory's mass matrix that implies a bound on the $\eta_V$ parameter for $d\ge 4$.  This bound agrees with the \emph{refined de Sitter conjecture} of \cite{Ooguri:2018wrx}: by virtue of the lemma in \cite{Andriot:2020wpp} (see footnote \ref{lemma}), one has in terms of the canonically normalised fields (\ref{canonfields})
\begin{equation}
   \eta_V:=\frac{\text{min}(\text{Eigen}({M^i}_{j}))}{V}\vert_{\text{crit}}\le -\frac{d}{2(d-2)}\simeq -\mathcal{O}(1)\, .
\end{equation}

Finally, let us comment that the AdS distance conjecture \cite{Lust:2019zwm} can also be verified.  This conjecture predicts the behaviour $m_{KK} \sim |\Lambda_{\text{AdS}}|^\alpha$ as the AdS curvature scale is taken to zero, with $\alpha \sim \mathcal{O}(1)$  and positive.  That is, the near flat-limit of AdS lies in the swampland as it is accompanied by an infinite tower of light KK states.  The strong version implies moreover that $\alpha \geq \frac12$ for non-supersymmetric AdS backgrounds and $\alpha=\frac12$ for supersymmetric AdS vacua.  As we discussed in subsection \ref{S:concreteegs}, our non-supersymmetric AdS$_7$ solution has no scale separation between the AdS curvature and the KK masses (from the $S^2$), with 
\begin{equation}
    \Lambda_{\text{AdS$_7$}}\equiv V_{\text{crit}}\sim \frac{M_s^2}{R^2}\sim m_{KK}^2\,,
\end{equation}
thus they saturate the swampland bound, with $\alpha=\frac12$.

\section{Conclusions}
\label{Conclusions}
We have investigated the possibility of de Sitter and anti de Sitter solutions in the context of Type II supersymmetry-breaking Scherk-Schwarz compactifications $\mathcal{M}_{1,d-1} \left(\times Y^m \right) \times \mathcal{T}_{\text{ss}}^n$, where $\mathcal{T}_{\text{ss}}^n$ is the Scherk-Schwarz torus and $Y^m$ a further possible compact factor.  This set up has several appealing features.  Firstly, the Scherk-Schwarz supersymmetry breaking leads to a 1-loop negative Casimir potential which breaks the strong energy condition $\tilde{T} \equiv -T^\mu_\mu + \frac{d}{D-2} T^M_M \geq 0$ that leads to the classic Maldacena-Nu\~nez de Sitter no-go theorem \cite{Maldacena:2000mw}.  Moreover, one can hope to use the interplay of very well-understood ingredients -- namely the toroidal Scherk-Schwarz 1-loop Casimir potential, the curvature of $Y^m$, and fluxes -- to stabilise the dilaton and volume moduli in a(n anti) de Sitter vacuum.  Among these, the fluxes can be introduced in such a way as to not source any RR tadpoles; there is thus no need for localised sources (and their associated smearing) or warping and flux numbers can be unbounded, which can be useful for parametric control.  Finally, concrete, backreacted solutions to the full 10d equations of motion and Bianchi identities can be obtained, whilst a lower-dimensional EFT analysis gives further general insights into the quality of any candidate solutions.

After reviewing the 1-loop Casimir potential and corresponding stress-energy tensor from Scherk-Schwarz compactifications (which assumes a flat background and vanishing fluxes and thus is a good approximation in our context in the large volume limit), we used the modified 10d, traced, traced-reversed Einstein equations and the dilaton equation to derive a series of no-go theorems for de Sitter solutions (dS no-gos \ref{signcas}-\ref{nogodim6}).  These can be alternatively read as giving necessary conditions for de Sitter solutions, under our assumptions of compactification geometry $\mathcal{M}_{1,d-1} \left(\times Y^m \right) \times \mathcal{T}_{\text{ss}}^n$ and tadpole-free fluxes.  In particular, for the single-factor compactifications, we found that de Sitter solutions require {\it{(1)}} a spacetime dimension $d \leq 7$; {\it{(2)}} a non-vanishing internal NSNS flux, $H_3$; {\it{(3)}} a non-vanishing internal RR flux, $F_q$, with $q<5$.  For the product compactifications, we found instead that de Sitter solutions require {\it{(1)}} a spacetime dimension $d \leq 5$; {\it{(2)}} the curvature of $Y^m$ to be positive, $\mathcal{R}_Y>0$; {\it{(3)}} an NSNS $H_3$ background flux, $H_3^{(0)}$, which fully threads the Scherk-Schwarz torus; {\it{(4)}} a RR background flux, $F_q^{(s_q)}$, with $s_q$ legs through $Y^m$, where $q<5-2s_q$.

Using these necessary conditions as a starting point, we then presented two explicit solutions to the full 10d equations of motion and Bianchi identities.   The first was a single-factor de Sitter solution, $dS_4 \times \mathcal{T}_{\text{ss}}^n$, which was, however,  seen to be outside the validity of our supergravity approximation and 1-loop Casimir computation, at least parametrically.  Indeed, since $g_s \sim R_{\text{ss}}^{2}$, it is impossible to obtain simultaneously -- and parametrically -- weak string coupling and large volume.  The second explicit example that we presented is an anti de Sitter solution, $AdS_7 \times S^2 \times S^1_{\text{ss}}$ (and we expect many other possible configurations); this solution can be easily put in a regime of parametric control using the unbounded flux numbers.  Both solutions are of course non-supersymmetric. 

Scale separation for any dS and AdS solutions can be excluded using the 10d equations of motion.  However, a lower-dimensional EFT description of the solutions can still be derived, from which more general insights into the parametric control of solutions and their perturbative stability can be drawn. 
Indeed, dimensional reduction on the product compactification $Y^m \times \mathcal{T}_{\text{ss}}^n$ leads to a $d$-dimensional EFT that encapsulates the dynamics of the universal moduli fields, the $d$-dimensional dilaton, $\phi_d$, and the respective volumes, $\chi$ and $\omega$ (the single-factor compactification can be retrieved by truncating the modulus $\chi$ and setting $m=0$).  We derived the EFT action, which includes an effective potential for the moduli from the curvature of $Y^m$, the flux terms, and the 1-loop Casimir potential. The subsequent extremisation conditions for this effective potential can be shown to be identical to the 10d dilaton and traced, traced-reversed Einstein equations that we used to derive the no-go theorems.  If, moreover, the choices of geometry and flux parameters required to find extrema can be shown to satisfy all the 10d equations of motion and Bianchi identities, then the EFT corresponds to a consistent truncation.  Consistency may also be argued for using symmetries, e.g. in the case of group manifolds such as torii or an $S^3$.  

For consistent truncations, the EFT provides a simple approach towards understanding the stability of corresponding 10d solutions; if an instability is found in the EFT, then the 10d solution is also unstable.  We then proved that {\it{any candidate de Sitter  solution has a perturbative instability amongst the universal moduli}}, and thus correspond to an unstable 10d solution.    Finally, the EFT -- in conjunction with the 10d necessary conditions for dS derived above --  also allows us to determine relations between the different control parameters for candidate de Sitter and anti de Sitter solutions.  Indeed, by considering the balance of different terms in the EFT potential, which descend from necessary ingredients in 10d, we were able to show that {\it{any de Sitter solution}} cannot simultaneously have parametrically weak string coupling and large volumes, since $g_s \sim R_{\text{ss}}^{2}$.  Our search for numerically controlled de Sitter solutions also failed.  On the other hand, we showed that parametrically controlled anti de Sitter solutions are easily obtained, by exploiting the unbounded flux numbers and a hierarchy of scales $R\gg R_{\text{ss}} \gg 1$.  Finally, we derived no-go theorems for slow-roll acceleration, and used our top-down string constructions to test various swampland conjectures in a non-supersymmetric setting.

To summarise, although supersymmetry breaking via Scherk-Schwarz toroidal compactifications a priori seemed a promising arena to obtain parametrically controlled 10d de Sitter solutions, we have proven -- conclusively -- that they are not possible under the assumptions of a geometry $\mathcal{M}_{1,d-1} \times Y^m \times \mathcal{T}_{\text{ss}}^n$ and tadpole-free flux backgrounds with no localised sources or warping.  No-go theorems have also been proven in 10d non-supersymmetric string theories \cite{Basile:2020mpt}.  It will be interesting to understand whether similar results are found in more refined Scherk-Schwarz compactifications involving Brane Supersymmetry Breaking or if these provide a path towards controlled de Sitter vacua in string theory \cite{MarcoKajal}.  On the other hand, we have shown that parametrically controlled non-supersymmetric AdS solutions are possible. Interestingly, the fluxes along the AdS spacetime can vanish in these solutions, and thus they would evade the WGC arguments against fully stable non-supersymmetric anti de Sitter used in \cite{Ooguri:2016pdq}.  However, the concrete example that we have presented also turns out to be perturbatively unstable (see \cite{Duff:1986hr, Gaiotto:2009mv, Lust:2009zb, Guarino:2020flh, Balaguer:2024cyb} for perturbatively stable non-supersymmetric anti de Sitter solutions).  Given this perturbative instability, and the absence of scale separation, our results are consistent with the conjecture \cite{Ooguri:2016pdq} that non-supersymmetric AdS holography does not exist in a quantum theory with a low energy description comprising Einstein gravity coupled to a finite number of matter fields, even if the AdS is not supported by flux.  Finally, returning to de Sitter solutions, whilst we have ruled out parametric control, we have not excluded numerical control.  It would be interesting to make a systematic exploration of the flux configuration space in order to establish if numerical control can be found.

\acknowledgments
It is our pleasure to thank David Andriot, Ivano Basile, Bruno Bento, Bruno De Luca, Joaquim Gomes, Ludwig Horer, Muthusamy Rajaguru, Alessandro Tomasiello, Dimitrios Tsimpis for helpful conversations.  The work of SLP is partially funded by STFC grant ST/X000699/1.  MS is funded by a H.G. Baggs Fellowship, University of Liverpool.

\appendix

\section{Scherk-Schwarz orbifold}\label{ssappendix}
\subsection{Modular forms}
 The generators of the $SL(2,\mathbb{Z})$ modular group are the $T: \tau\rightarrow \tau+1$ and $S: \tau\rightarrow -\frac{1}{\tau}$ modular transformations. In terms of $q:=e^{2\pi i\tau}$, we define  the \emph{Dedekind eta function} 
\begin{equation}
\eta(\tau):=q^{\frac{1}{24}}\prod_{n=1}^\infty(1-q^n)\, ,
\end{equation}
and the \emph{Jacobi theta functions} 
\begin{equation}
    \theta\left[\begin{smallmatrix}
        \alpha\\
        \\
        \beta
    \end{smallmatrix}\right](z,\tau)=\sum_{n=-\infty}^{+\infty}e^{2\pi i(n+\alpha)(z+\beta)}q^{(n+\alpha)^2/2}\, .
\end{equation}
Under the modular transformations $T$ and $S$ these functions transform as
\begin{subequations}
\begin{align}
    \eta(\tau+1)&=e^{\pi i/12}\,\eta(\tau)\\
    \theta\left[\begin{smallmatrix}
        \alpha\\
        \\
        \beta
    \end{smallmatrix}\right](z,\tau+1)&=e^{-i\pi\alpha(\alpha-1)}\theta\left[\begin{smallmatrix}
        \alpha\\
        \\
        \beta+\alpha-1/2
    \end{smallmatrix}\right](z,\tau)\\[0.3em]
      \eta(-1/\tau)&=\sqrt{-i\tau}\,\eta(\tau)\\
      \theta\left[\begin{smallmatrix}
        \alpha\\
        \\
        \beta
    \end{smallmatrix}\right](z,-1/\tau)&=\sqrt{-i\tau}e^{2i\pi\alpha\beta+i\pi z^2/\tau}\theta\left[\begin{smallmatrix}
        \beta\\
        \\
        -\alpha
    \end{smallmatrix}\right](z,\tau) \,.
\end{align}    
\end{subequations}
For our purposes we consider the following four Jacobi theta functions\footnote{Notice that $\theta_1(\tau)=0$ identically.}
\begin{equation}
\theta_1(\tau):=\theta \left[\begin{smallmatrix}
        1/2\\
        \\
        1/2
    \end{smallmatrix}\right](0,\tau)\, ,\quad \theta_2(\tau):=\theta \left[\begin{smallmatrix}
        1/2\\
        \\
        0
    \end{smallmatrix}\right](0,\tau)\, ,\quad \theta_3(\tau):=\theta \left[\begin{smallmatrix}
        0\\
        \\
        0
    \end{smallmatrix}\right](0,\tau)\, ,\quad \theta_4(\tau):=\theta \left[\begin{smallmatrix}
        0\\
        \\
        1/2
    \end{smallmatrix}\right](0,\tau)
\end{equation}
from whose fourth powers  we define the \emph{$SO(8)$ characters}
\begin{equation}
    \begin{split}
 V_8&=\frac{\theta_3^4-\theta_4^4}{2\eta^4}\\
 O_8&=\frac{\theta_3^4+\theta_4^4}{2\eta^4}\\
 S_8&=\frac{\theta_2^4+\theta_1^4}{2\eta^4}\\
 C_8&=\frac{\theta_2^4-\theta_1^4}{2\eta^4} \,.\\
    \end{split}
\end{equation}
They obey the \emph{Jacobi identity} 
\begin{equation}
    V_8=S_8=C_8\, ,
\end{equation}
and under the modular transformations transform as
\begin{equation}
   \frac{1}{\eta^8} \begin{pmatrix}
        O_8\\
        V_8\\
        S_8\\
        C_8
    \end{pmatrix}(\tau+1)=\frac{1}{\eta^8}\begin{pmatrix}
       - O_8\\
        V_8\\
        S_8\\
        C_8
    \end{pmatrix}\,, \quad  \frac{1}{\eta^8} \begin{pmatrix}
        O_8\\
        V_8\\
        S_8\\
        C_8
    \end{pmatrix}\left(-\frac{1}{\tau}\right)=\frac{1}{2}\frac{1}{(-i\tau)^4}\frac{1}{\eta^8}\begin{pmatrix}
        O_8+V_8+S_8+S_8\\
        O_8+V_8-S_8-C_8\\
        O_8-V_8+S_8-C_8\\
        O_8-V_8-S_8+C_8
    \end{pmatrix}
\end{equation}
In the following we will need the $q$-expansions of the characters divided by the eighth power of the $\eta$ function
\begin{equation}\label{qexp}
\begin{split}
    \frac{V_8}{\eta^8}[q]=\frac{S_8}{\eta^8}[q]= \frac{C_8}{\eta^8}[q]=\frac{\theta_2^4}{2\eta^{12}}[q]&=8\sum_{k=0}^{\infty} d_k q^{k}\,,\quad d_0=1\, ,\\
    \frac{O_8}{\eta^8}[q]&=\sum_{k=0}^\infty c_k q^{k-\frac{1}{2}}\, ,\quad c_0=1\, .\\
\end{split}    
\end{equation}
It will also be useful to define $\Gamma_0^1[2]$ and $\Gamma_0[2]$  congruence subgroups of 
$SL(2,\mathbb{Z})$, 
\begin{equation}
\begin{split}
    \Gamma_0 (2) &= \left\{ \begingroup\renewcommand{\arraystretch}{0.9}\scalebox{0.8}{$\begin{pmatrix}
        a & b \\
        c & d
    \end{pmatrix}$}\endgroup
    \in \mathrm{SL}(2,\mathbb{Z}) : \, c \equiv 0 \pmod{2} \right\},\\
    \Gamma_0^1 (2) &= \left\{ \begingroup\renewcommand{\arraystretch}{0.9}\scalebox{0.8}{$\begin{pmatrix}
        a & b \\
        c & d
    \end{pmatrix}$}\endgroup \in \mathrm{SL}(2,\mathbb{Z}) : \, c \equiv 0 \pmod{2} \, \wedge \, d \equiv 1 \pmod{2} \right\}.
 \end{split} 
\end{equation}

\subsection{Torus partition function}
In string theory, the 1-loop effective potential for oriented closed strings in arbitrary spacetime dimension $D$ can be computed by integrating the \emph{torus partition function} $\mathcal{T}$ over the $SL(2,\mathbb{Z})$ fundamental domain $\mathcal{F}$
\begin{equation}
    \Lambda_{\text{1-loop}}=-\frac{1}{2}\frac{M_s^{D}}{(2\pi)^D}\int_{\mathcal{F}}\frac{d^2\tau}{\tau_2^2}\mathcal{T}\, ,
\end{equation}
where
\begin{equation}
    \mathcal{F}:=\{\tau\in \mathbb{C}:\,\ab \text{Re}\,\tau\ab\le \frac{1}{2},\,\text{Im}\tau>0,\,\ab \tau\ab\ge 1\}\,,
\end{equation}
with also $\tau=\tau_1 + i \tau_2$, and schematically
\begin{equation}\label{torusampli}
\mathcal{T}=\text{Str}\,q^{L_0-\frac{c}{24}}\,\bar{q}^{\bar{L}_0-\frac{\bar{c}}{24}}\, .
\end{equation}
In the above formula $L_0$, $\bar{L}_0$ are the  Virasoro operators, $c$, $\bar{c}$ the central charges of the left and right-moving sectors and the Supertrace is computed over the  Hilbert space of the theory, including  the momentum space.

The torus partition function for 10d type IIB string theory is obtained from the above formula once the GSO projection $P_{\text{GSO}}=\frac{1}{4}(1-(-1)^F)(1-(-1)^{\bar{F}})$ is inserted into the trace and reads
\begin{equation}\label{tiib10d}
\mathcal{T}_{\text{IIB}}=\frac{1}{\tau_2^4}\left\ab\frac{V_8-S_8}{\eta^8}\right\ab^2\, .
\end{equation}
Here $(\sqrt{\tau_2}\,\eta\,\bar{\eta})^8$ is the contribution of the bosons associated to the 8 transverse dimensions, where a factor $\sqrt{\tau_2}$ for each transverse dimension arises from the Gaussian integration over its continuous momentum, and  the $V_8$ and $S_8$ characters encodes respectively the GSO projected traces over the NS and R sectors. 

The latter straightforwardly generalises to IIB toroidal compactifications. If  $d=10-D$ dimensions are compactified on a torus $T^d$, with metric $G_{IJ}$ and its inverse $G^{-1}_{IJ}=G^{IJ}$, the associated internal momenta are now discrete and given in the left and right-moving sectors by
\begin{equation}    p_{I}^L=m_I+G_{IJ}n_J\,,\quad p_{I}^R=m_I-G_{IJ}n_J\, ,
\end{equation}
where $m_I,n_I\in \mathbb{Z}$ are respectively the quantised KK momentum and winding numbers along the compact direction $I=D,\dots,9$. As such, the trace over the  compact momenta  no longer contributes with a factor $\sqrt{\tau_2}$ for each compact dimension from the Gaussian integration but rather with a discrete sum over the lattice $\Lambda_{\vec{m},\vec{n}}$. Therefore, performing  in (\ref{tiib10d}) the replacement
\begin{equation}
    {\tau_2}^\frac{10-D}{2}\rightarrow \sum_{\vec{m},\vec{n}}\Lambda_{\vec{m},\vec{n}}(\tau):=q^{\frac{1}{4}P^L_IG^{IJ}P^L_J}\bar{q}^{\frac{1}{4}P^R_IG^{IJ}P^R_J}\, ,
\end{equation}
we obtain the torus partition function for IIB compactified on a $d$-dimensional torus down to $D$ dimensions
\begin{equation}\label{toruscomp}
\mathcal{T}_{\mathrm{IIB}_{\mathrm{T^d}}}=\frac{1}{(\sqrt{\tau_2})^{D-2}}\left\ab\frac{V_8-S_8}{\eta^8}\right\ab^2\Lambda_{\vec{m},\vec{n}}(\tau)\, .
\end{equation}

In general, once the characters and the lattice are expressed in terms of their $q,\bar{q}$-series (\ref{qexp}), the torus partition function will contain terms of the form
\begin{equation}
    \int d^2\tau\Lambda_{\vec{m},\vec{n}}[\tau,\bar{\tau}]q[\tau]^k\bar{q}[\bar{\tau}]^l= \int d^2\tau e^{2\pi i\tau_1(k-l-m_In^I)}e^{-2\pi\tau_2(k+l+\frac{1}{2}m_IG^{IJ}m_J+n_IG_{IJ}n_J)}\, .
\end{equation}
Physical states obey to the level-matching condition, which is obtained by perfoming the $\tau_1$-integration and  reads
\begin{equation}
    k-l=m_In^I\, .
\end{equation}
The squared-masses of the states are given by
\begin{equation}
    M^2=2(k+l)+m_IG^{IJ}m_J+n_IG_{IJ}n_J\, .
\end{equation}

\subsection{Supersymmetry-breaking}
Supersymmetry-breaking \`a la Scherk-Schwarz can be implemented in string theory at the worldsheet level in a modular invariant way by means of a freely acting orbifold.

In its simplest realisation, we compactify IIB on a circle $S^1$ of radius $R_{\text{ss}}$ (units of $\sqrt{\alpha'}$) and project the resulting spectrum with the $\mathbb{Z}_2$ generator $g=(-1)^F\delta_{KK}$, where $F$ is the total spacetime fermion number and $\delta_{KK}$ is the freely-acting \emph{momentum shift}
\begin{equation}\label{momentumshift}
    \delta_{KK}:\,\left(x_L^9,x_R^9\right)\rightarrow\left( x^9_L+\frac{1}{2}\pi R_{\text{ss}}\sqrt{\alpha'}, x^9_R+\frac{1}{2}\pi R_{\text{ss}}\sqrt{\alpha'}\right)\, ,
\end{equation}
with $x^9=x_L^9+x_R^9$ the coordinate of the compact dimension. At the level of the torus partition function, this amounts to inserting the projector $P_g=(1+g)/2$
into the trace (\ref{torusampli}). Taking into account the IIB torus partition function for the $S^1$ compactification ($D=9$ in (\ref{toruscomp})), the tentative orbifolded torus partition function would be
\begin{equation}
\begin{split}
  \mathcal{T}^\star_{\mathrm{IIB}_{\mathrm{S^1}}/g}&=\frac{1}{2}\mathcal{T}_{\mathrm{IIB}_{\mathrm{S^1}}}+\frac{1}{2}\text{Str}_{\text{IIB}}\,g\,q^{L_0-\frac{c}{24}}\,\bar{q}^{\bar{L}_0-\frac{\bar{c}}{24}}\\
  &=\frac{1}{2\tau_2^{7/2}}\left(\left\ab \frac{V_8-S_8}{\eta^8}\right\ab^2\Lambda_{m,n}+\left\ab \frac{V_8+S_8}{\eta^8}\right\ab^2(-1)^m\Lambda_{m,n}\right)\, ,
 \end{split} 
\end{equation}
where in the second term we clearly see the action of the orbifold $g$ as $(-1)^FS_8=-S_8$ and $\delta_{KK}\Lambda_{m,n}=(-1)^m\Lambda_{m,n}$. 
Modular invariance, however, constrains us to consider the full orbit of the second term under the modular transformations
\begin{equation}\label{chain}
    \stackrel{T}{\circlearrowright}\ab V_8+S_8\ab^2(-1)\Lambda_{m,n}\stackrel{S}{\leftrightarrows}\ab O_8-C_8\ab^2\Lambda_{m,n+\frac{1}{2}}\stackrel{T}{\leftrightarrows}\ab O_8+C_8\ab^2(-1)\Lambda_{m,n+\frac{1}{2}} \stackrel{S}{\circlearrowleft}
\end{equation}
such that eventually the modular invariant torus partition function must contain  the characters $O_8$, $C_8$ associated to the``reversed'' GSO projection
\begin{equation}\label{s1kkshift}
\begin{split}
  \mathcal{T}_{\mathrm{IIB}_{\mathrm{S^1}}/g}=\frac{1}{2\tau_2^{7/2}}\biggl(&\left\ab \frac{V_8-S_8}{\eta^8}\right\ab^2\Lambda_{m,n}+\left\ab \frac{V_8+S_8}{\eta^8}\right\ab^2(-1)^m\Lambda_{m,n}\\
  &+\left\ab \frac{O_8-C_8}{\eta^8}\right\ab^2\Lambda_{m,n+\frac{1}{2}}+\left\ab \frac{O_8+C_8}{\eta^8}\right\ab^2(-1)^m\Lambda_{m,n+\frac{1}{2}}\biggr)\, .
 \end{split} 
\end{equation}
The states described by the first (second) line of characters are usually referred to as untwisted (twisted) states.
At this point, it is customary to switch from the orbifold basis to the Scherk-Schwarz basis by rescaling $R_\text{ss}\rightarrow 2R{_\text{ss}}$, such that the torus partition function becomes familiar with the field theory expectation
\begin{equation}
\begin{split}
  \mathcal{T}_{\mathrm{IIB}_{\mathrm{S^1}}/g}=&(V_8\bar{V}_8+S_8\bar{S}_8)\Lambda_{m,2n}+(V_8\bar{S}_8+S_8\bar{V}_8)\Lambda_{m+\frac{1}{2},2n}\\
  +&(O_8\bar{O}_8+C_8\bar{C}_8)\Lambda_{m,2n+1}+(O_8\bar{C}_8+C_8\bar{O}_8)\Lambda_{m+\frac{1}{2},2n+1}\, .
 \end{split} 
\end{equation} 
All the fermions now have half-integer KK number and hence their zero-modes acquire mass, signalling supersymmetry-breaking. The gravitino mass in the untwisted NS-R sector sets the scale of supersymmetry-breaking
\begin{equation}
    M_{\text{gra}}=\frac{M_s}{2\,R{_\text{ss}}}\sim M_{KK}\, ,
\end{equation}
which vanishes in the supersymmetric limit $R_\text{ss}\rightarrow \infty$.
As a universal feature of standard Scherk-Schwarz deformations in string theory, we then observe that the untwisted states  come with even winding numbers while the twisted states come with odd winding numbers.
The lowest-lying twisted state from the $O_8$ character therefore has a non-vanishing winding number and its mass reads
\begin{equation}\label{tac}
         M^2_{\text{tac}}=-\frac{2}{\alpha'}+\frac{R_{\text{ss}}^2}{\alpha'}\, ,
\end{equation}
 hence it is tachyonic for $R{_\text{ss}}<\sqrt{2}\equiv R_H$. However for large values of the radius $R{_\text{ss}}\gg 1$ this tachyon is lifted and the perturbative description still holds. In this limit, the twisted states are  very massive and can be safely integrated out from the EFT.

Supersymmetry-breaking by momentum shifts can also be realised in IIA. For oriented closed strings only, they are completely equivalent by T-duality to a winding shift $\delta_w$ in the type IIB language 
\begin{equation}
    \delta_{w}:\,(x_L^9,x_R^9)\rightarrow\left( x^9_L+\frac{1}{2}\pi\frac{\sqrt{\alpha'}}{R{_\text{ss}}}, x^9_R-\frac{1}{2}\pi\frac{\sqrt{\alpha'}}{R{_\text{ss}}}\right)\, ,
\end{equation}
with  $\delta_{w}\Lambda_{m,n}=(-1)^n\Lambda_{m,n}$. From the IIB point of view, the tachyonic region (\ref{tac}) is now pushed to the large $R_\text{ss}$ limit while the spectrum is free of tachyon for small $R_{\text{ss}}<1$ and supersymmetry is recovered in the $R_{\text{ss}}\rightarrow 0$ limit, where  the IIA picture is well defined. 

The supersymmetry-breaking mechanism just described for an $S^1$ compactification immediately generalises to $d$-dimensional toroidal $T^d$ compactifications. The momentum deformation we adopt in this work is the one of a $\delta_{KK}$ shift performed along every compact direction, such that $\delta_{KK}\Lambda_{\vec{m},\vec{n}}=(-1)^{\vec{m}\cdot\vec{\epsilon}}\Lambda_{\vec{m},\vec{n}}$, where $\vec{m}=(m_1,\dots, m_d)$, $\vec{n}=(n_1,\dots ,n_d)$ are the vectors of KK and winding numbers respectively and $\vec{\epsilon}=(1,1,\dots,1)_d$ is the $d$-dimensional unit-vector. The corresponding torus partition function is a direct generalisation of the $S^1$ case (\ref{s1kkshift}) and reads
\begin{equation}\label{torustd}
\begin{split}
  \mathcal{T}_{\mathrm{IIB}_{\mathrm{T^d}}/g}=\frac{1}{2\tau_2^\frac{D-2}{2}}\biggl(&\left\ab \frac{V_8-S_8}{\eta^8}\right\ab^2\Lambda_{\vec{m},\vec{n}}+\left\ab \frac{V_8+S_8}{\eta^8}\right\ab^2(-1)^{\vec{m}\cdot \vec{\epsilon}}\Lambda_{\vec{m},\vec{n}}\\
  &+\left\ab \frac{O_8-C_8}{\eta^8}\right\ab^2\Lambda_{\vec{m},\vec{n}+\frac{1}{2}\vec{\epsilon}}+\left\ab \frac{O_8+C_8}{\eta^8}\right\ab^2(-1)^{\vec{m}\cdot\vec{\epsilon}}\Lambda_{\vec{m},\vec{n}+\frac{1}{2}\vec{\epsilon}}\biggr)\, ,
 \end{split} 
\end{equation}
which can be rearranged as
\begin{equation}
  \begin{split}
  \mathcal{T}_{\mathrm{IIB}_{\mathrm{T^d}}/g}=&(V_8\bar{V}_8+S_8\bar{S}_8)\Lambda_{\vec{m},\vec{n}}\vert_{\vec{m}\cdot\vec{\epsilon}\in 2\mathbb{Z}}+(V_8\bar{S}_8+S_8\bar{V}_8)\Lambda_{\vec{m},\vec{n}}\vert_{\vec{m}\cdot\vec{\epsilon}\in 2\mathbb{Z}+1}\\
  +&(O_8\bar{O}_8+C_8\bar{C}_8)\Lambda_{\vec{m},\vec{n}+\frac{1}{2}\vec{\epsilon}}\vert_{\vec{m}\cdot\vec{\epsilon}\in 2\mathbb{Z}}+(O_8\bar{C}_8+C_8\bar{O}_8)\Lambda_{\vec{m},\vec{n}+\frac{1}{2}\vec{\epsilon}}\vert_{\vec{m}\cdot\vec{\epsilon}\in 2\mathbb{Z}+1}\, .
 \end{split} 
\end{equation}
At the point in moduli space where the internal torus is factorisable, i.e $T^d=S^1(R_1)\times\dots\times S^1(R_d)$, the lattice factorises as well $\Lambda_{\vec{m},\vec{n}}=\Lambda_{m_1,n_1}(R_1)\cdot\cdot\cdot \Lambda_{m_d,n_d}(R_d)$. After the customary rescalings $R_i\rightarrow 2R_i$, the mass of the would-be tachyon from the NS-NS twisted sector is
\begin{equation}
M^2_{\text{tac}}=-\frac{2}{\alpha'}+\frac{1}{\alpha'}\sum_{i=1}^d R_i^2\,,
\end{equation}
and again the spectrum is tachyon-free in the large $R_i\gg 1$ limit. 
\subsection{One-loop potential}\label{one-loopotA}
To compute the 1-loop potential  from the Scherk-Schwarz torus partition function (\ref{torustd}) one needs to unfold the fundamental domain \cite{PhysRevD.36.1184},\cite{Kiritsis:1997hf},\cite{Trapletti:2002uk} in the $G_{ij}\gg 1$ limit where the spectrum is tachyon-free and the partition function stays finite.

Considering the modular transformation chain (\ref{chain}), we can rewrite the modular integral up to the $\frac{M_s^D}{2(2\pi)^{D}}$ prefactor as
\begin{equation}\label{lambdatd}
    \Lambda_{\text{1-loop}}=-\frac{1}{2}\int_\mathcal{F}\frac{d^2\tau}{\tau_2^2}\frac{\sqrt{\text{det}(G)}}{\tau_2^4}\sum_{g\in \cal{G}}g\,\circ \left[\left\ab\frac{V_8+S_8}{\eta^8}\right\ab^2\, \sum_{{m}^i,{n}^i\in\mathbb{Z}}e^{-\frac{\pi}{4\tau_2}G_{ij}\left[((2m^i+1)+2\,n^i\tau)((2m^j+1)+2\,n^j\bar{\tau})\right]} \right]\, ,
\end{equation}
where $\mathcal{G}=\{1,S,TS\}$, a subgroup of $SL(2,\mathbb{Z})$, and  we have performed a Poisson resummation over $m^i$. 
The lattice can be further recast in the following form (see appendix H of \cite{Kiritsis:1997hf}, for $B_{ij}=0$)
\begin{equation}\label{latticeA}
    \sum_{{m}^i,{n}^i\in\mathbb{Z}}e^{-\frac{\pi}{4\tau_2}G_{ij}\left[((2m^i+1)+2\,n^i\tau)((2m^j+1)+2\,n^j\bar{\tau})\right]}=\sum_{A\in \text{Mat}_{d\times 2}}e^{-\frac{\pi\,T_2}{2}-\frac{\pi\,T_2}{4\tau_2U_2}\ab \tau-\bar{U}\ab^2}\, ,
\end{equation}
where 
\begin{equation}\label{A}
    A^T=\begin{pmatrix}
        2n_1&2n_2&\cdot\cdot\cdot&2n_d\\
        2m_1+1&2m_2+1&\cdot\cdot\cdot&2m_d+1\, 
    \end{pmatrix}\, ,
\end{equation}
\begin{equation}
\begin{split}
T_2&=\sqrt{\hat{G}_{11}\hat{G}_{22}-\hat{G}^2_{12}}\, ,\\
U&=\frac{1}{\hat{G}_{11}}\left(-\hat{G}_{12}+i\sqrt{\hat{G}_{11}\hat{G}_{22}-\hat{G}^2_{12}}\right)\, ,
\end{split}
\end{equation}
with $T_2$ and $U$,  respectively, the imaginary part of the K\"ahler form and the complex structure induced from the  pullback $\hat{G}$ 
\begin{equation}
    \hat{G}_{IJ}:= M^i_IG_{ij}M^j_J\,,\quad M=A^T.
\end{equation}
Exploiting $SL(2,\mathbb{Z})$ transformations, we can therefore consider the orbits of the congruence subgroup $\Gamma_0^1(2)$ to replace the sum in (\ref{latticeA}) over $2\times d$ matrices of the form (\ref{A})  with a sum over representatives  of such orbits. There are two orbits, a degenerate orbit with representatives of the form
\begin{equation}
    A_0^T=\begin{pmatrix}
        0&0&\cdot\cdot\cdot&0\\
        2m_1+1&2m_2+1&\cdot\cdot\cdot&2m_d+1\, 
    \end{pmatrix}\,,\quad 
\end{equation}
and a non-degenerate one with representatives
\begin{equation}\label{nondeg}
    A_1^T=\begin{pmatrix}
        2n_1&2n_2&\cdot\cdot\cdot &2n_k&0&\cdot\cdot\cdot&0\\
        2m_1+1&2m_2+1&\cdot\cdot\cdot&2m_k+1&2m_{k+1}+1&\cdot\cdot\cdot&2m_d+1\end{pmatrix}\,,\quad  1\le k<d,\quad 2n_k>2m_k+1>0\, .
\end{equation}
Notice that $A^T_0$ is invariant under $T$ modular transformations, implying that it actually generates orbits of $\Gamma_0^1(2)/T$.
The rest of the torus partition function can be rewritten as a Dedekind eta quotient
\begin{equation}
    \frac{1}{\tau_2^4}\left\ab \frac{V_8+S_8}{\eta^8}[\tau]\right\ab^2=\frac{1}{\tau_2^4}\left\ab \frac{\theta_2^4}{\eta^{12}}[\tau]\right\ab^2=\frac{256}{\tau_2^4}\left\ab \frac{\eta^8(2\tau)}{\eta^{16}(\tau)}\right\ab^2\, ,
\end{equation}
which 
is invariant under the congruence subgroup $\Gamma_0(2)\supset \Gamma_0^1(2)$ \cite{sussman2017rademacherseriesetaquotients}. Hence, we can rewrite in total
\begin{equation}
\begin{split}
    \Lambda_{\text{1-loop}}=-\frac{1}{2}\int_\mathcal{F}\frac{d^2\tau}{\tau_2^2}\frac{\sqrt{\text{det}(G)}}{\tau_2^4}&\sum_{g\in \mathcal{G}}g\,\circ \biggl\{\sum_{\gamma\in \Gamma_0^1(2)/T}\gamma\circ\left[\, \sum_{A_0\in \text{Mat}_{d\times 2}}e^{-\frac{\pi\,T_2}{2}-\frac{\pi\,T_2}{4\tau_2U_2}\ab \tau-\bar{U}\ab^2}\left\ab\frac{\theta_2^4}{\eta^{12}}\right\ab^2 \right]\\
    &+\sum_{\gamma\in \Gamma_0^1(2)}\gamma\circ\left[\, \sum_{A_1\in \text{Mat}_{d\times 2}}e^{-\frac{\pi\,T_2}{2}-\frac{\pi\,T_2}{4\tau_2U_2}\ab \tau-\bar{U}\ab^2}\left\ab\frac{\theta_2^4}{\eta^{12}}\right\ab^2 \right]\biggr\}\, .
 \end{split}   
\end{equation}
Performing a change of variable and using that $(1+S+TS)\circ \Gamma_0^1(2)=SL(2,\mathbb{Z})$, we have
\begin{equation}
    \sum_{g\in \mathcal{G},\gamma\in \Gamma_0^1(2)/T}g\circ \gamma(\mathcal{F})=\mathcal{S}\,,\quad   \sum_{g\in \mathcal{G},\gamma\in \Gamma_0^1[2]}g\circ\gamma(\mathcal{F})=2\,\mathcal{\mathbb{C}^+}\,,
\end{equation}
implying that the fundamental domain unfolds, respectively,  into the strip $\mathcal{S}$, defined as the image of the fundamental domain $\mathcal{F}$ under the modular group, modulo unitarity translations on the real axis, i.e
\begin{equation}
    \mathcal{S}:=\{\tau \in \mathbb{C}:\,\ab \text{Re}\,\tau\ab\le \frac{1}{2},\,\text{Im}\tau>0\}=\frac{SL(2,\mathbb{Z})}{T}(\mathcal{F})\, ,
\end{equation}
and the double cover of the upper-half complex plane. 

The contribution from the degenerate orbit, $\Lambda_{\text{1-loop},0}$, is thus found by setting to zero all the winding numbers, inserting the $q$-expansion for the characters $V_8,S_8$ (\ref{qexp})  and computing the integral
\begin{equation}
\begin{split}
    \Lambda_{\text{1-loop},0}&=-2\cdot64\,\sqrt{\text{det}(G)}\,\int_{-\frac{1}{2}}^{\frac{1}{2}} d\tau_1 \int_{0}^{+\infty}\frac{d\tau_2}{\tau_2^6} \sum_{k,\bar{k}\in \mathbb{N}}\sum_{m^i\in \mathbb{Z}} c_{k}c_{\bar{k}}e^{-\frac{\pi}{4\tau_2}G_{ij}(2m^i+1)(2m^j+1)-2\pi\tau_2(k+\bar{k})}e^{2\pi i\tau_1(k-\bar{k})}\,.
 \end{split}   
\end{equation}
The $\tau_1$-integration simply imposes the level-matching condition $k=\bar{k}$. Computing the $\tau_2$ integral using 
\begin{equation}\label{tau2int}
    \int_0^{+\infty} \frac{d\tau_2}{\tau_2^{1+\lambda}}e^{-c\,\tau_2-\frac{b}{\tau_2}}=\frac{2}{b^\lambda}(c\,b)^\frac{\lambda}{2}K_\lambda\left(2\sqrt{c\,b}\right)=\frac{1}{b^\lambda}\,\Gamma(\lambda)\,\mathcal{H}_\lambda\left(\sqrt{c\,b}\right)\, ,
\end{equation}
where $K_\lambda$ is a modified Bessel function of the second kind and $\mathcal{H}_\lambda$ is defined through the second equality above, we find eventually
\begin{equation}
     \Lambda_{\text{1-loop},0}=-2\cdot 64 \cdot\frac{4^5}{\pi^5}\, \Gamma(5)\sqrt{\text{det}(G)}\sum_{k\in \mathbb{N}}\sum_{m^i\in \mathbb{Z}}c_k^2\,\frac{\mathcal{H}_5\left[\pi\sqrt{k}\sqrt{(2m^i+1)G_{ij}(2m^j+1)}\right]}{[(2m^i+1)G_{ij}(2m^j+1)]^5}\, .
\end{equation}
For the non-degenerate orbit we need to compute 
\begin{equation}
\begin{split}
   \Lambda_{\text{1-loop},1}=-2\times\frac{1}{2}\cdot4\cdot64\,\sqrt{\text{det}(G)}\int_{-\infty}^{+\infty} d\tau_1 \int_{0}^{+\infty}\frac{d\tau_2}{\tau_2^6}\sum_{k,\bar{k}\in \mathbb{N}}\sum'_{m^i,n^i\in \mathbb{Z}} c_{k}c_{\bar{k}}&e^{-\frac{\pi}{4\tau_2}G_{ij}(2m^i+1+2n^i\tau)(2m^j+1+2n^j\bar{\tau})}\\
   &\times e^{2\pi i\tau_1(k-\bar{k})-2\pi\tau_2(k+\bar{k})},
 \end{split}   
\end{equation}
where the sums $\sum'_{m^i,n^i\in \mathbb{Z}}$ are subjected to the constraints in (\ref{nondeg}). In this case the $\tau_1$-integral is Gaussian. We are then left with the $\tau_2$-integral, again of the form (\ref{tau2int})
\begin{equation}
\begin{split}
    \Lambda_{\text{1-loop},1}&=-4\cdot64\sqrt{\text{det}(G)}\int_0^{+\infty}\frac{d\tau_2}{\tau_2^{11/2}}\sum_{k,\bar{k}\in \mathbb{N}}\sum'_{m^i,n^i\in \mathbb{Z}} \frac{c_{k}c_{\bar{k}}}{\sqrt{n^iG_{ij}n^j}}e^{-\frac{\pi}{4\tau_2}\left((2m^i+1)G_{ij}(2m^j+1)-\frac{((2m^i+1)G_{ij}n^j)^2}{n^iG_{ij}n^j}\right)}\\
    &\qquad\qquad\qquad\times e^{-\pi\tau_2\left(n^iG_{ij}n^j+2(k+\bar{k})+\frac{(k-\bar{k})^2}{n^iG_{ij}n^j}\right)}e^{-\pi i(k-\bar{k})\frac{n^iG_{ij}(2m^j+1)}{n^iG_{ij}n^j}}\\
    &=64\frac{4^{11/2}}{\pi^{9/2}}\Gamma\left(\frac{9}{2}\right)\sqrt{\text{det}(G)}\sum_{k,\bar{k}\in \mathbb{N}}\sum'_{m^i,n^i\in \mathbb{Z}} \frac{c_{k}c_{\bar{k}}}{\sqrt{n^iG_{ij}n^j}}\,\frac{cos\left(\pi(k-\bar{k})\frac{n^iG_{ij}(2m^j+1)}{n^iG_{ij}n^j}\right)}{\left((2m^i+1)G_{ij}(2m^i+1)-\frac{(n^iG_{ij}(2m^j+1))^2}{n^iG_{ij}n_j}\right)^9}\\
    & \times \mathcal{H}_{\frac{9}{2}}\left[\frac{\pi}{2}\left(n^iG_{ij}n^j+2(k+\bar{k})+\frac{(k-\bar{k})^2}{n^iG_{ij}n^j}\right)^\frac{1}{2}\left((2m^i+1)G_{ij}(2m^j+1)-\frac{(n^iG_{ij}(2m^j+1))^2}{n^iG_{ij}n_j}\right)^\frac{1}{2}\right]
    \end{split}
\end{equation}
The total 1-loop effective potential is given by $\Lambda_{\text{1-loop}}=\Lambda_{\text{1-loop},0}+\Lambda_{\text{1-loop},1}$. We then note that the argument of $\mathcal{H}_5$ is $\mathcal{O}\left(\sqrt{G_{ij}}\right)$,  except when $k=0$ in which case it vanishes, while the argument of $\mathcal{H}_{\frac{9}{2}}$ is always $\mathcal{O}(G_{ij})$. According to the limiting behaviours of the $\mathcal{H}_\lambda$ function for small and large arguments respectively
\begin{equation}\label{limitingh}
    \mathcal{H}_\lambda(z)\sim \frac{\sqrt{\pi}}{\Gamma(\lambda)}\,z^{\lambda-\frac{1}{2}}\,e^{-2\,z}\,\quad \text{if}\,\, z\gg 1\,,\qquad \mathcal{H}_\lambda(z)= 1-\frac{z^2}{\lambda-1}+\mathcal{O}(z^4)\,\quad \text{if}\, \,\ab z\ab\gg 1\, ,
\end{equation}
this means that in the $G_{ij}\gg 1$ regime the leading  order contribution to the 1-loop potential comes exclusively from the $k=0$ term in the degenerate orbit, which represents the contribution from the KK modes of  the massless string states $(k=0)$ propagating along the susy-breaking directions. All the other states, e.g.  massive string states and KK modes thereof, states with non trivial winding masses and non level-matched states, are very massive  and therefore yield  exponentially suppressed contributions. Using (\ref{limitingh}) we can hence write the 1-loop potential in the $G_{ij}\gg 1$ limit, now reinserting the $\frac{M_s^D}{2(2\pi)^D}$ prefactor, as
\begin{equation}\label{lambda1loop}
    \Lambda_{\text{1-loop}}= -64\,\frac{2^{10-D}\Gamma(5)}{\pi^{D+5}}\,M_s^D \sqrt{\text{det}(G)}\sum_{m^i}\frac{1}{[(2m^i+1)G_{ij}(2m^j+1)]^5}+\mathcal{O}\left(e^{-\text{min}(G_{ij})}\right)\, .
\end{equation}
The associated 10d stress-energy tensor is 
\begin{equation} \label{TMNgentor}
    T_{MN}=-64 \frac{2^{10-D}\Gamma(5)}{\pi^{D+5}}M_s^{10}\sum_{\vec{m}}\frac{1}{[(2m^i+1)G_{ij}(2m^j+1)]^5}\left(-G_{MN}+10\frac{(2m^k+1)(2m^l+1)G_{kM}G_{lN}}{[(2m^i+1)G_{ij}(2m^j+1)]}\right)
\end{equation}
and is readily checked to be traceless. 

To finish, if the Scherk-Schwarz torus is assumed to be factorisable and with a single radius $R_{\text{ss}}$, in the limit $R_{\text{ss}}\gg 1$ and after performing the rescaling $R_{\text{ss}}\rightarrow 2 R_{\text{ss}}$ to pass from the orbifold basis to the Scherk-Schwarz/EFT basis, the 1-loop effective potential (\ref{lambda1loop}) can be recast into the final form
\begin{equation}\label{vssA}
    V_{\text{ss}}=(n_f^0-n_b^0)\,\xi_n\, \left(\frac{M_s}{R_{\text{ss}}}\right)^D+\mathcal{O}\left(e^{-R_{\text{ss}}}\right)\, ,
\end{equation}
where $n_f^0-n_b^0=-64$ and 
\begin{equation} \label{xin}
\xi_n:= \frac{2^{2n-10}}{\pi^{15-n}}\Gamma(5)\sum_{m_i=-\infty}^{+\infty}\frac{1}{[(2m_1+1)^2+\dots +(2m_n+1)^2]^5}\simeq 
\frac{3\cdot\,2^{3n-7}}{\pi^{15-n}}\,\frac{1}{n^5}\,.
\end{equation}

\section{10d supergravity}
\label{app:10dsugra}
The starting point of our construction is the type IIB and (massive) IIA 10d supergravities. The bosonic spectrum of these theories is described by a universal NSNS sector, populated by the 10d-dilaton $\Phi$, the Kalb-Ramond 2-form $B_2$ and the 10d metric $G_{MN}$, and a RR sector of $p$-form $C_p$ potentials, respectively  $C_1,C_3$ in IIA ($p$ odd) and $C_0,C_2,C_4$ in IIB ($p$ even). At the tree-level and in the string-frame, these bulk fields are described by following  string-frame 10d action
\begin{equation}\label{IIaction}
    S_{\text{tree}}^{\text{IIB/A}}=S_{\text{NS}}+S^{\text{IIB/A}}_{\text{RR}}+S^{\text{IIB/A}}_{\text{CS}}
\end{equation}
where the universal NSNS part is 
\begin{equation}
    S_{\text{NS}}=\frac{1}{2\kappa^2_{10}}\int d^{10}x\sqrt{-G_{10}}\,e^{-2\Phi}\left(R_{10}+4(\partial \Phi)^2-\frac{1}{2}\ab H_3\ab^2\right)\, ,
\end{equation}
 and the RR actions are, respectively,
\begin{equation}
    \begin{split}
        S^{\text{IIB}}_{\text{RR}}&=-\frac{1}{4\kappa^2_{10}}\int d^{10}x\sqrt{-G_{10}}\left(\,\ab F_1\ab^2+\ab F_3\ab^2+\frac{1}{2}\ab F_5\ab^2\right)\, ,\\
          S^{\text{IIA}}_{\text{RR}}&=-\frac{1}{4\kappa^2_{10}}\int d^{10}x\sqrt{-G_{10}}\left(\,\ab F_0\ab^2+\ab F_2\ab^2+\ab F_4\ab^2\right)\, .
    \end{split}
\end{equation}
We do not need to specify the Chern-Simons term $S_{\text{CS}}$.  

  In our conventions
\begin{equation}
    2\kappa^2_{10}\equiv\frac{\ell_s^8}{2\pi}\,,\quad \ell_s:=2\pi\sqrt{\alpha'}\, .
\end{equation}
The field strengths appearing in the actions are the NSNS 3-form $H_3=dB_2$, the (odd) RR forms $F_1=dC_0$, $F_3=dC_2-C_0\wedge H_3$ and $F_5=dC_4-\frac{1}{2}C_2\wedge H_3+\frac{1}{2}B_2\wedge F_3$ in IIB and the (even) RR forms  $F_2=dC_1+B_2\wedge F_0$ and $F_4=dC_3-C_1\wedge H_3+\frac{1}{2}F_0 B_2\wedge B_2$ in IIA. The Romans mass $F_0$ is a constant present in the massive deformation of IIA supergravity. $F_5$ obeys a self-duality condition $F_5=\star_{10}F_5$ that must be imposed on-shell. For a 10d  $p$-form $A_p$ we denote
\begin{equation}
    \ab A_p\ab^2=\frac{1}{p!}\,G^{M_1N_1}\cdot\cdot\cdot G^{M_pN_p}{A}_{p\,M_1\dots M_p}A_{p\,N_1\dots N_p}\,.
\end{equation}
The Hodge-star operator $\star_D$ in dimension $D$ acts with the Levi-Civita symbol $\epsilon_{0\dots D-1}=1$ as
\begin{equation}
    \star_D (dx^{m_1}\wedge\dots \wedge dx^{m_p})=\frac{\sqrt{\text{det}(g_D)}}{(D-p)!}g^{m_1n_1}\cdot\cdot\cdot g^{m_pn_p}\,\epsilon_{n_1\dots n_p r_{p+1}\dots r_{D}}\,dx^{m_{p+1}}\wedge\dots \wedge dx^{r_D}\, .
\end{equation}

The equations of motion stemming from this action are the flux equations of motion, the dilaton equation of motion and the Einstein equations; we should also consider the Bianchi identities.  Starting from the fluxes, in absence of localized sources the IIA and IIB  Bianchi identities for the RR fluxes reads
\begin{equation}\label{fluxbi}
\begin{split}
dF_0&=0\,,\\
    dF_{2}&-F_0\,H_3=0\,,\\
    d F_4&-H_3\wedge F_2=0\, ;\\
    dF_1&=0\,\\
    \quad dF_{3}&-H_3\wedge F_1=0\,,\\
     d F_5&=d(\star_{10}F_5)=H_3\wedge F_3\,,
\end{split}
\end{equation}
and the RR flux equations of motion are given by
\begin{equation}\label{fluxeom}
\begin{split}
d(\star_{10}F_1)&+H_3\wedge\star_{10}F_3=0\,,\\ d(\star_{10}F_3)&+H_3\wedge\star_{10}F_5=0\, ,\\
    d(\star_{10}F_2)&+H_3\wedge\star_{10}F_4=0\,,\\
    d(\star_{10}F_4)&+H_3\wedge F_4=0\,.
\end{split}
\end{equation}
Lastly, the  equation of motion and the Bianchi identity for the NSNS 3-form $H_3$ are
\begin{equation}\label{h3eom}
    \begin{split}
  &d(e^{-2\Phi}\star_{10}H_3)-\sum_{1\le q\le 4}F_{q-1}\wedge \star_{10}F_{q+1}-\frac{1}{2}F_{4}\wedge F_{4}=0\\
   \text{and \quad}&dH_3=0\,.
   \end{split} 
\end{equation}
The dilaton equation is
\begin{equation}\label{10ddilatoneq}
    2R_{10}=\ab H_3\ab^2-8(\Box\Phi-\ab\partial\Phi\ab^2)\,.
\end{equation}
The type II supergravity stress energy tensors are
\begin{equation}
    \begin{split}
\,\kappa^2_{10}\,e^{2\Phi}T^{\text{SUGRA,IIA}}_{MN}
=&\frac{1}{4}{H_3}_{MPQ}{{H_3}_N}^{PQ}+\frac{1}{2}e^{2\Phi}\left({F_2}_{MP}{{F_2}_N}^P+\frac{1}{3!}{F_4}_{MPQR}{{F_4}_N}^{PQR}\right)\\
    &-\frac{g_{MN}}{4}\left(\ab H_3\ab^2+e^{2\Phi}\left(\ab F_0\ab^2+\ab F_2\ab^2+\ab F_4\ab^2\right)\right)\\
&-2\nabla_M\partial_N\Phi+2g_{MN}\left(\Box \Phi-\ab\partial\Phi\ab^2\right)\\
\kappa^2_{10}\,e^{2\Phi}T^{\text{SUGRA,IIB}}_{MN}
=&\frac{1}{4}{H_3}_{MPQ}{{H_3}_N}^{PQ}+\frac{e^{2\Phi}}{2}\left({F_1}_M{F_1}_N+\frac{1}{2}{F_3}_{MPQ}{{F_3}_N}^{PQ}+\frac{1}{2\cdot 4!}{F_5}_{MPQRS}{{F_5}_{N}}^{PQRS}\right)\\
    &-\frac{g_{MN}}{4}\left(\ab H_3\ab^2+e^{2\Phi}\left(\ab F_1\ab^2+\ab F_3\ab^2\right)\right)\\
&-2\nabla_M\partial_N\Phi+2g_{MN}\left(\Box \Phi-\ab\partial\Phi\ab^2\right)\, .
    \end{split} \label{10dTsugra}
\end{equation}
The trace-reversed Einstein equations $\mathcal{R}_{MN}
=\kappa^2_{10} \,e^{2\Phi}\left({T^{\text{SUGRA,IIA/B}}_{MN}}-\frac{1}{8}G_{MN}T^{\text{SUGRA,IIA/B}}_{(10)}\right)$ read, respectively, for IIA and IIB 

\begin{align}
    \mathcal{R}_{MN}
=\,&\frac{1}{4}{H_3}_{MPQ}{{H_3}_N}^{PQ}+\frac{e^{2\Phi}}{2}\left({F_2}_{MP}{{F_2}_{N}}^P+\frac{1}{3!}{F_4}_{MPQR}{{F_4}_N}^{PQR}\right)\nonumber\\
    &-\frac{g_{MN}}{8}\left(\ab H_3\ab^2-\frac{1}{2}e^{2\Phi} \left(\ab F_0\ab^2-\ab F_2\ab^2-3\ab F_4\ab^2\right)
+2\Box\Phi-4\ab\partial\Phi\ab^2\right)-2\nabla_M\partial_N\Phi\nonumber\, ,\\
\mathcal{R}_{MN}=\,&\frac{1}{4}{H_3}_{MPQ}{{H_3}_N}^{PQ}+\frac{e^{2\Phi}}{2}\left({F_1}_M{F_1}_N+\frac{1}{2}{F_3}_{MPQ}{{F_3}_N}^{PQ}+\frac{1}{2\cdot 4!}{F_5}_{MPQRS}{{F_5}_{N}}^{PQRS}\right)\nonumber\\
    &-\frac{g_{MN}}{8}\left(\ab H_3\ab^2+e^{2\Phi}\ab F_3\ab^2
+2\Box\Phi-4\ab\partial\Phi\ab^2\right)-2\nabla_M\partial_N\Phi    \,.  \label{10dEinstein}
\end{align}

\section{Some useful flux identities}
\label{app:comps}
The following identities are useful in the derivation of the flux terms in the Einstein traces (\ref{mcurv}) and (\ref{intcurv}) in subsubsection \ref{10secan}.
For the internal fluxes with $q\le 5$, we have 
\begin{equation}
    \begin{split}
    \delta^{ab}{F^{\text{int}}_q}_{aM_1\dots M_{q-1}}{{F^{\text{int}}_q}_b}^{M_1\dots M_{q-1}}&=(q-1)!\,\sum_{s_q}s_q\ab F_q^{(s_q)}\ab^2_\Int,\\
     \delta^{ij}{F^{\text{int}}_q}_{iM_1\dots M_{q-1}}{{F^{\text{int}}_q}_j}^{M_1\dots M_{q-1}}&=(q-1)!\,\sum_{s_q}(q-s_q)\ab F_q^{(s_q)}\ab^2_\Int\, ,\\
    \end{split}
\end{equation}
with $F_5^{\text{int}}$ and $F_6^{\text{int}}$ also getting extra contributions from their dual $F_5^{\text{ext}}$ and $F_{4}^{\text{ext}}$
external fluxes  that can be present in $d=3,4$:
\begin{equation}
    \begin{split}
        \delta^{ab}{F^{\text{ext}}_5}_{aM_1\dots M_{4}}{{F^{\text{ext}}_{5}}_b}^{M_1\dots M_{4}}&=-4!\,\sum_{s_q=0}^5(m-s_5)\ab F_5^{(s_5)}\ab^2_\Int\\
      \delta^{ij}{F^{\text{ext}}_5}_{iM_1\dots M_{4}}{{F^{\text{ext}}_5}_j}^{M_1\dots M_{4}}&=-4!\,\sum_{s_q=0}^5((\delta_{d,4}+2\delta_{d,3})+s_5-m)\ab F_5^{(s_5)}\ab^2_\Int \\
     \delta^{ab} {F^{\text{ext}}_4}_{aM_1\dots M_{3}}{{F^{\text{ext}}_4}_b}^{M_1\dots M_{3}}&=-3!\,\sum_{s_6=0}^5(m-s_6)\ab F_6^{(s_6)}\ab^2_\Int\\
      \delta^{ij}{F^{\text{ext}}_4}_{iM_1\dots M_{3}}{{F^{\text{ext}}_4}_j}^{iM_1\dots M_{3}}&=-3!\,\sum_{s_6=0}^5(\delta_{d,3}+s_6-m)\ab F_6^{(s_6)}\ab^2_\Int\, .
    \end{split}
\end{equation}


 \bibliographystyle{JHEP}
 \bibliography{biblio.bib}


\end{document}